\documentclass[backref, twocolumn, breaklinks, colorlinks]{aastex62}   
\hypersetup{linkcolor=blue,citecolor=blue,filecolor=cyan,urlcolor=magenta}

\usepackage{enumitem}
\usepackage{mathtools}
\usepackage{graphicx}
\usepackage{amssymb}
\usepackage{amsmath}
\usepackage{ulem}
\usepackage{epstopdf}
\usepackage{color}

\DeclareMathOperator\erf{erf}
\newcommand{\sign}{\text{sgn}}

\begin{document}
\sloppy

\title{{Pressure Balance and Intrabinary Shock Stability in Rotation-Powered State Redback and Transitional Millisecond Pulsar Binary Systems}}

\author{Zorawar Wadiasingh}
\affil{Astrophysical Science Division, NASA Goddard Space Flight Center, Greenbelt, MD 20771, USA}
\affil{Centre for Space Research, North-West University, Private Bag X6001, Potchefstroom 2520, South Africa}

\author{Christo Venter}
\affil{Centre for Space Research, North-West University, Private Bag X6001, Potchefstroom 2520, South Africa}

\author{Alice K. Harding}
\affil{Astrophysical Science Division, NASA Goddard Space Flight Center, Greenbelt, MD 20771, USA}

\author{Markus B\"ottcher}
\affil{Centre for Space Research, North-West University, Private Bag X6001, Potchefstroom 2520, South Africa}

\author{Patrick Kilian}
\affil{Centre for Space Research, North-West University, Private Bag X6001, Potchefstroom 2520, South Africa}

\shorttitle{Pressure Balance Scenarios of tMSPs}

\begin{abstract} 
A number of low-mass millisecond pulsar (MSP) binaries in their rotation-powered state exhibit double-peaked X-ray orbital modulation centered at inferior pulsar conjunction. This state, which has been known to persist for years, has recently been interpreted as emission from a shock that enshrouds the pulsar. However, the pressure balance for such a configuration is a crucial unresolved issue. We consider two scenarios for pressure balance: a companion magnetosphere and stellar mass loss with gas dominance. It is found that the magnetospheric scenario requires several kilogauss poloidal fields for isobaric surfaces to enshroud the MSP as well as for the magnetosphere to remain stable if there is significant mass loss. For the gas-dominated scenario, it is necessary that the companion wind loses angular momentum prolifically as an advection or heating-dominated flow. Thermal bremsstrahlung cooling in the flow may be observable as a UV to soft X-ray component independent of orbital phase if the mass rate is high. We formulate the general requirements for shock stability against gravitational influences in the pulsar rotation-powered state for the gas-dominated scenario. We explore stabilizing mechanisms, principally irradiation feedback, which anticipates correlated shock emission and companion variability and predicts $ F_\gamma/F_{\rm X} \lesssim 14$ for the ratio of pulsar magnetospheric $\gamma$-ray to total shock soft-to-hard X-ray fluxes. This stability criterion implies an unbroken extension of X-ray power-law emission to hundreds of keV for some systems. We explore observational discriminants between the gas-dominated and magnetospheric scenarios, motivating contemporaneous radio through $\gamma$-ray monitoring of these systems. 
\end{abstract} 


\section{Introduction}

The current decade has ushered in a new era for rotation-powered millisecond pulsars (MSPs) with radio, X-ray, and optical followup of unidentified \emph{Fermi} Large Area Telescope sources yielding over 30 pulsar binaries\footnote{https://confluence.slac.stanford.edu/display/GLAMCOG/\newline~Public+List+of+LAT-Detected+Gamma-Ray+Pulsars} in the Field. Recent population synthesis studies suggest the existence of order $\sim 10^4$ MSPs in the Galactic bulge and several hundred in each nearby globular cluster \citep{2018arXiv180611215G}. Moreover, the known population of MSPs is expected to surge enormously in the coming decade with the advent of the Square Kilometer Array \citep{2015aska.confE..40K} and FAST \citep{2009A&A...505..919S,2011IJMPD..20..989N}.

The subset of rotation-powered MSPs in detached binaries with low-mass companions are classified based on inferred companion mass $m_{\rm c}$, the ``black widows" (BWs) with $m_{\rm c} \lesssim 0.05 M_\odot$ \citep{1988Natur.333..237F} and ``redbacks" (RBs) with $m_{\rm c} \gtrsim 0.1 M_\odot$ \citep{2011AIPC.1357..127R}. For both classes, orbits are circularized with periods $P_b$ less than a day and inferred separation $a \sim 10^{11}$ cm. The tidally-locked companions in RBs are bloated (compared a main-sequence star of similar mass) close to the Roche limit and are anisotropically heated, often with distinct day and night halves. In the standard evolutionary scenario of recycled MSPs, a low-mass X-ray binary accretion-powered spin-up phase \citep[LMXB; ][]{1982Natur.300..728A} precedes a pulsar rotation-powered spin-down state. The LMXB state can attain Eddington-scale X-ray luminosities $L_{\rm X} \lesssim10^{38}$ erg s$^{-1}$, precipitated by Roche lobe overflow (RLOF) and the formation of an accretion disk. A subset of neutron star LMXBs are the accreting millisecond X-ray pulsars \citep[AMXP, cf.][for a review]{2012arXiv1206.2727P} where the disk is truncated at the Alfv\'{e}n radius $r_{\cal A}\sim 10^7 \left[\dot{m}_{\rm g}/(10^{15} \, \rm g \, s^{-1})\right]^{-2/7}$~cm from the pulsar, the point at which MSP magnetospheric magnetic pressure balances accretion pressure of mass rate $\dot{m}_{\rm g}$. For these millisecond spin periods $P_{\rm MSP}$, $r_{\cal A}$  is similar to the small pulsar light cylinder radius $r_{\rm LC} = c P_{\rm MSP}/(2 \pi) \sim$ few $\times 10^7$ cm $\sim 10^{-4} a$. Propellor states may exist if the disk inner Keplerian speed is smaller than the MSP corotation speed inside $r_{\rm LC}$ \citep[e.g,][]{1975A&A....39..185I}, i.e., when $r_{\cal A}$ is larger than the corotation Keplerian radius. 

Accretion-derived irradiation of the companion influencing mass loss has been invoked in long-term gigayear evolutionary models of LMXBs for their formation \citep{1988Natur.334..225K, 1989ApJ...336..507R,1989ApJ...343..292R,1991Natur.351...39T,2004A&A...423..281B}, following the first suggestion of such ``autoregulation" in the context of $\alpha$-disks by \cite{1973A&A....24..337S}. That is, emission from the accretion disk irradiating the companion bootstraps the mass loss and the accretion power. Such binary evolutionary tracks make simplifying assumptions about the poorly understood radiatively-driven winds or mass loss from the companion and the spectral energy distribution (SED) of the accretion luminosity. More recent work focused on BW and RB formation has found that irradiation feedback induces mass transfer cycles between rotation and accretion-powered states in the late-term evolution, with periods on the order of $10^6$ years, and predicts that RB companions should slightly under-fill their Roche lobe, the ``quasi-Roche lobe overflow" model \citep[qRLOF: ][]{2014ApJ...786L...7B,2015ApJ...798...44B}. The existence of transitional systems and these models then imply conditions where the wind of the companion may be evaporative and supersonic rather than the more conventional higher-mass-rate subsonic RLOF, and also regimes intermediate between these two limits. Evaporative qRLOF would entail a wind from the high-energy tail of the Maxwellian in the photosphere or corona, that is sufficient to escape the low potential barrier of a companion slightly underfilling its Roche lobe.  

Recently, some systems have been observed to transition between accretion and rotation-powered states \citep{2009Sci...324.1411A,2013Natur.501..517P,2014ApJ...789...40B,2015ApJ...800L..12R}, persisting for years $\tau_{\rm p} \sim 10^8$~s in one state preceding or following a transition. The transition itself may occur on a short timescale, shorter than a few weeks as sampled by typical observational cadences \citep{2014MNRAS.441.1825B}. In the disk phase, these transitional systems may exhibit complex X-ray phenomenology interpreted as propeller, sub-luminous or active/passive disk, or accretion states occasionally with coherent MSP spin pulsations similar to an AMXP \citep{2014ApJ...795...72L,2015ApJ...807...33P,2015MNRAS.449L..26P,2017ApJ...851L..34P}. The X-ray persistent luminosities of transitional systems in disk states can be relatively low, $L_{\rm X} \lesssim 10^{35}$ erg s$^{-1}$ suggesting relatively low mass rates, $|\dot{m}_{\rm c}| \lesssim 10^{16}$ g s$^{-1}$ accreted from the low-mass companion for a standard $10\%$ radiative efficiency. It has been advanced by \cite{2015MNRAS.447.3034H} that some subset of very faint X-ray binaries may be such transitional systems or AMXPs. 

It is unknown if the companion fills its Roche lobe during the disk states in transitional MSPs. For many RBs in the rotation-powered state, the companions are known to be close but not quite filling their Roche lobe \citep[e.g.,][]{2015MNRAS.451.3468M,2016ApJ...816...74B}. However, small changes in the Roche filling factor or radiatively-driven wind physics may dramatically alter the mass loss rate in the transition between qRLOF and RLOF. The donor star need not significantly change its radius on short timescales associated with rotation-powered and accretion state changes if it is already nearly filling its Roche lobe. 

In the pulsar state, radio or $\gamma$-ray magnetospheric pulsations of the MSP are often observable. We are not aware of any optical evidence of disks in the pulsar state, in contrast to the accretion state \cite[e.g.,][]{2013ATel.5514....1H}. In this rotation-powered pulsar state, many BWs \citep{2012ApJ...760...92H,2014ApJ...783...69G} and RBs \cite[e.g.,][]{2015arXiv150207208R} exhibit persistent nonthermal and hard X-ray emission with photon indices typically $1-1.5$ (see Table~\ref{RBenergetics}). Thermal X-ray emission, besides that ascribed to the MSP polar caps \citep{2011ApJ...742...97B}, is absent. For J1023+0038 during its rotation-powered phase, no break in the power law was detected with {\textit{NuSTAR}} up to at least $50$ keV \citep{2014ApJ...791...77T}. Similarly, J1723--2837 \citep{2017ApJ...839..130K} and J2129--0429 \citep{2018ApJ...861...89A,2018arXiv180601312K} also exhibit no spectral cut-off at {\it{NuSTAR}} energies. Because of the rising spectra in a $\nu F_\nu$ representation, the highest energies of the power laws dominate the energetics of this component. 

Moreover, about nine systems exhibit orbital modulation of the persistent X-ray emission \citep[see Table 1 of ][]{2017ApJ...839...80W}, with $L_{\rm X} \sim 10^{32}-10^{33}$ erg s$^{-1}$, including emission in the {\it{NuSTAR}} band and even in some systems with inclinations far from edge-on \citep[e.g., J1023+0038,][]{2010ApJ...722...88A,2011ApJ...742...97B,2014ApJ...791...77T}. A possible tenth system omitted in \cite{2017ApJ...839...80W} is J1740--5340 \citep{2003ApJ...584L..13F,2010ApJ...709..241B}. These X-ray orbital phase-folded light curves are often double-peaked, with a local minimum either near pulsar superior or inferior conjunction, which we denote SCDP (companion between pulsar and Earth)  or ICDP (pulsar in front), respectively. Such ICDP orbital modulation is especially striking in J2129--0429 \citep{2015arXiv150207208R,2018ApJ...861...89A,2018arXiv180601312K} and J1227--4853 \citep{2015MNRAS.454.2190D}, among others. In the disk state of J1023+0038, \cite{2015ApJ...806..148B} find no evidence of orbital modulation in either the low, high or flaring modes of the X-ray emission. Therefore the orbital modulation in the pulsar state is of a qualitatively different origin than X-rays in the disk state, and is, by definition causally associated with the stellar companion and its orbital timescale.

\floattable
\begin{deluxetable*}{lcccccccccccccc}
\tabletypesize{\tiny}
\setlength{\tabcolsep}{0.02in}
\tablecaption{X-ray and $\gamma$-ray Energetics of RBs with Pulsar Inferior Conjunction Double Peaked Phase Centering (ICDP) \label{RBenergetics}}
\tablecolumns{14}
\tablehead{ \colhead{ Name } & \colhead{$F_{\gamma}$\tablenotemark{a}} & \colhead{$F_{\rm Xs}$\tablenotemark{b}} & \colhead{$F_{\rm Xsh}$\tablenotemark{c}} & \colhead{$\Gamma_{\rm Xs}$} & \colhead{$F_{\gamma}/F_{\rm Xs}$} & \colhead{$F_{\gamma}/F_{\rm Xsh}$}  & \colhead{$\varepsilon_{\rm min,cut}$\tablenotemark{d}} &  \colhead{$\log_{10} \dot{E}_{\rm SD}$\tablenotemark{e}} &  \colhead{D\tablenotemark{f}} & \colhead{$\log_{10}  L_\gamma$\tablenotemark{g}} & \colhead{$ \log_{10}  L_{\rm Xs}$\tablenotemark{g}} & \colhead{$\log_{10}  L_{\rm Xsh}$\tablenotemark{g}} & \colhead{Refs.} }
\startdata
J1023+0038\tablenotemark{*} & $0.50 \pm 0.09 \pm 0.05$ & $3.8 \pm 0.1$ & $33 \pm 4$ & $1.19 \pm 0.03$ & 13 & 1.5 & -- & 35.1 & $1.37$ & 33.1 & 31.9 & 32.9 & (1)   \\
J1227--4853\tablenotemark{*} & $1.79 \pm 0.16 \pm 0.17$ & $4.6 \pm 0.1$ & -- & $1.2 \pm 0.04$ & 39 & -- & $\boldsymbol{35}\,(25-48)$ & 35.1 & $1.8$ & 33.8 & 32.2 & -- & (2)   \\
J1723--2837 & $0.83 \pm 0.23 \pm 0.4$ & $18.7 \pm 0.2$ & $96 \pm 5$ & $1.13 \pm 0.02$ & 4 & 0.9 & -- & 34.8 & $0.74$ &  32.7 & 32.1 & 32.8 & (3)   \\
J2129--0429 & $1.10 \pm 0.08 \pm 0.03$ & $2.1 \pm 0.23$ & $15 \pm 2$ & $1.13 \pm 0.08$ & 52 & 7 & -- & 34.7 &  $1.83$ & 33.6 & 31.9 & 32.8 &  (4)   \\
J2215+5135 & $1.33 \pm 0.09 \pm 0.06$ & $1 \pm 0.3$ & -- & $1.4 \pm 0.2$ & 130 & -- & $\boldsymbol{260}\,(80-2700)$ &  34.9 & $3.0$ & 34.2 & 32.0 & -- & (5)   \\
J2339--0533 & $5.05 \pm 0.19 \pm 0.53$ & $2.50 \pm 0.15$ & -- & $1.32 \pm 0.08$ & 200 & -- & $\boldsymbol{420}\,(220-900)$ & 34.5 & $1.1$ & 33.9 & 31.6 & -- & (6)  
\enddata
\tablenotetext{a}{Phase-averaged \emph{Fermi}-LAT flux, adopted from Table 1 of \cite{2017ApJ...836...68T}, in units of $10^{-11}$ erg cm$^{-2}$ s$^{-1}$.}
\tablenotetext{b}{Classical soft X-ray band fluxes in units of $10^{-13}$ erg cm$^{-2}$ s$^{-1}$. }
\tablenotetext{c}{{\it NuSTAR} soft-hard $3-79$ keV fluxes in units of $10^{-13}$ erg cm$^{-2}$ s$^{-1}$.}
\tablenotetext{d}{In keV, to satisfy the inequality Eq.~(\ref{etalimit}). Uncertainties in $F_{\gamma}$, $F_{\rm Xs}$ and $\Gamma_{\rm Xs}$ are accounted in the parenthetical range.}
\tablenotetext{e}{$\dot{E}_{\rm SD}$ in erg s$^{-1}$ with a fiducial moment of inertia of $1.3\times 10^{45}$ g cm$^{2}$.}
\tablenotetext{f}{Distance in kiloparsec, adopted from the ATNF catalog.}
\tablenotetext{g}{Approximate isotropic $\gamma$-ray luminosities computed in erg s$^{-1}$, neglecting uncertainties in distance and energy flux.}
\tablenotetext{*}{All quantities quoted are for the rotation-powered radio MSP epochs.}
\tablerefs{ATNF Catalog: \cite{2005AJ....129.1993M}
(1) \cite{2010ApJ...722...88A,2011ApJ...742...97B,2012ApJ...756L..25D,2014ApJ...791...77T}
(2) \cite{2014ApJ...789...40B,2015ApJ...800L..12R}
(3) \cite{2014ApJ...781....6B,2017ApJ...839..130K}
(4) \cite{2015arXiv150207208R,2018ApJ...861...89A,2018arXiv180601312K}
(5) \cite{2014ApJ...783...69G}
(6) \cite{2015ApJ...802...84Y}
}
\end{deluxetable*}

Some scenarios for the persistent emission can be ruled out owing to its energetics. It may be shown that orbital energy extraction by any mechanism would yield too short an inspiral timescale if it is to entirely power the persistent X-ray emission, in disagreement with much smaller known $\dot{P}_b$ constraints. The nonthermal ICDP modulated component is also difficult to explain as originating internally from the stellar companion. The minimum putative energy output is roughly $L_{\rm X} \tau_{\rm p} \gtrsim 10^{41}$ erg, of order $10^{-7}$ of the stellar gravitational binding energy. Even with $100\%$ conversion efficiency, this is much larger than the magnetic reservoir $R_{\rm c}^3 B_*^2 \sim 10^{37}-10^{39}$ erg for kilogauss magnetic fields attainable in convection-dominated low-mass stars. 
Moreover, if the ICDP emission is powered by persistent companion-intrinsic magnetic activity, then it is unclear why there is no evidence for it in the disk state low-mode of J1023+0038 \citep{2015ApJ...806..148B} where it may contribute $\sim 15-25\%$ of the observed flux in the soft X-ray band \citep{2010ApJ...722...88A,2011ApJ...742...97B}. It is also unknown how such magnetically-powered activity would naturally yield persistent nonthermal high-energy ICDP modulation across many sources. Therefore, magnetic activity may only account for more transient phenomena. Furthermore, if there is no shock, the solid angle fraction of the pulsar wind intercepted by the companion is of order $10^{-2}$ -- this would then demand untenably large conversion efficiency of $\dot{E}_{\rm SD}$ at around $100\%$ into the hard X-rays at the companion. Such pure wind conversion is also contradicted by relatively cool optically-derived photospheric temperatures $T <10^4$ K for RB companions. Therefore, the source of the persistent nonthermal X-ray emission is not proximate to the companion photosphere. 

Phase-resolved X-ray hardness ratios of orbital modulation in ICDP systems exhibit a harder-when-brighter phenomenology \citep[e.g.,][]{2010ApJ...722...88A,2011ApJ...742...97B,2015MNRAS.454.2190D,2015ApJ...801L..27H}. This as well as {\it{NuSTAR}} power laws beyond 30 keV, rule out absorption as an origin of the orbital modulation (however, absorption may play a role in stability of a shock in some scenarios, as we explore in this paper). Moreover, absorption or obscuration of a putative disk emission by the companion or its wind neither yields double peaks nor modulation at the correct inferior-conjunction phasing. Some RBs in the pulsar state also exhibit large radio MSP eclipse fractions, $>50\%$ of the orbit at low frequencies for RBs J1023+0038 and J2215+5135. Crucially, the pulsar is largely uneclipsed around pulsar inferior conjunction in eclipsing RBs \citep[e.g.,][]{2009Sci...324.1411A,2013arXiv1311.5161A,2016MNRAS.459.2681B,2018JPhCS.956a2004M}. These large orbitally phase-dependent eclipses, and lack thereof at pulsar inferior conjunction, are another feature unexpected if there exists a disk outside $r_{\rm LC}$. Similar to the orbital modulation in the persistent X-ray emission, orbital-phase-dependent eclipses are causally associated with the companion. Note that radio eclipsing BWs such as B1957+20 appear to have qualitatively different eclipses than RBs such as J1023+0038; eclipses in BWs like B1957+20 are much shorter in duration and more regular \citep[and similarly in BW J1810+1744,][]{2018MNRAS.476.1968P} around superior conjunction of the MSP. 

An interpretation of the above phenomenology of ICDP systems is that an intrabinary pulsar termination shock accelerates electrons that rapidly cool principally via synchrotron radiation, similar to that surmised in the SCDP-type system BW B1957+20 \citep{1990ApJ...358..561H,1993ApJ...403..249A,2012ApJ...760...92H} but with a shock curving around the MSP in ICDP systems. The shock subtends a solid angle from the pulsar much larger than the companion, with the total power budget is constrained by the pulsar spin-down power $\dot{E}_{\rm SD} \sim 10^{34} - 10^{35}$ erg s$^{-1}$. It may naturally account for the X-ray energetics and large orbital radio eclipses. The double-peaked light curves are putatively generated by Doppler-boosting of the synchrotron emission in a mildly relativistic flow along the termination shock with the X-ray double-peak modulation phase centering providing a discriminant of the shock orientation \citep{2016arXiv160603518R,2017ApJ...839...80W}. The Doppler beaming may arise from MHD-like fast magnetosonic flows \citep[e.g.,][]{2002AstL...28..373B,2002MNRAS.336L..53B,2004MNRAS.349..779K}, or kinetically by anisotropic particle distributions along a shear layer \citep{2013ApJ...766L..19L,2017ApJ...847...90L}. The level of orbital modulation ascribed to Doppler-boosted shock emission is related to the binary inclination, among other factors, inhibiting identification of low-inclination systems if the MSP is not detectable in some epochs due to transitions to disk states. ICDP systems may thus involve an intrabinary shock oriented around the pulsar and well inside the pulsar Roche lobe. In this paper, we consider the energetics and stability of this configuration. The putative termination shock stagnation point is past the $L_1$ point of the companion within the MSP Roche lobe, yet still well outside the pulsar light cylinder since the pulsar mechanisms are operational. 

We note that since the shock radiative power cannot exceed $\dot{E}_{\rm SD} \sim 10^{35}$ erg s$^{-1}$, this limits the power-law extension to a few or tens of MeV. Physically, the maximum photon energy $\epsilon_{\rm max}$ of the power-law extension is approximately set by the unknown maximum shock-accelerated electron/positron Lorentz factor $\gamma_{\rm e, max}$, ignoring Doppler factors of order unity. Since the shock emission is putatively synchrotron emission, the maximum energy is roughly $m_e c^2 \epsilon_{\rm max} \sim \gamma_{\rm e, max}^2 (B/B_{\rm cr})$ where $B_{\rm cr} = 4.413 \times 10^{13}$ G and $B$ is the post-shock magnetic field, expected to be on the order of a few Gauss \citep[Eq. (5) in][]{2017ApJ...839...80W}. Then, $ \epsilon_{\rm max} \gtrsim 1 (\equiv 0.511 \, \rm MeV)$ provided that $ \gamma_{\rm e, max} \gtrsim 10^7$. Such high Lorentz factors are generally accepted in pulsar wind termination shocks \citep[e.g.,][]{1996ApJ...457..253D, 2017hsn..book.2159S, 2017JPhCS.932a2050K}. 

In this paper, we suggest two scenarios for pressure balance for a putative shock curved around the pulsar. These are delineated as asymptotic limits of the plasma parameter $\beta =  8 \pi n_l k_b T_l/B_l^2$ where $n_l$, $T_l$ and $B_l$ are the local plasma rest frame number density, temperature and magnetic field in the plasma arising from the companion, respectively. The companion plasma will not be everywhere either magnetically or gas-dominated, but for practicality we consider these two limits. Clearly, in the disk state $\beta \gtrsim 1$ but it is unclear if gas dominance persists in the pulsar state. 

We first consider a strong companion magnetosphere in \S\ref{comppressure} (hereafter Scenario $\beta \ll 1$) where $\beta \ll 1$ everywhere prior to the shock and the companion wind gas pressure play no dynamically important role for the shock. We find that a sufficiently strong companion magnetic dipole moment will yield a curved quasi-hemispherical termination shock around the MSP, regardless of the orientation of the putative dipole moment, even for anisotropic pulsar winds, provided that the MSP spin and orbital axis are parallel. Scenario $\beta \ll 1$ is also stable insofar as the companion magnetosphere is stable and the companion mass loss rate is low. 

The other limit, Scenario $\beta \gg 1$, is examined in \S\ref{gravsec}, where mass loss from the companion provides the pressure balance for the shock formation, i.e., the magnetic field plays no dynamically important role. The formation of a shock instead of a disk imposes constraints on the character of the companion wind and mass loss, and energetic arguments suggest the wind is gravitationally captured by the MSP in this scenario, conceivably by an advection-dominated-accretion-flow-like solution (ADAF). The ADAF premise and its observational consequences are examined in \S\ref{ADAF}. However, in isolation, such a shock-ADAF configuration is inherently unstable to gravitational influences on dynamical timescales \citep{2001ApJ...560L..71B}, therefore stabilizing mechanisms ought to exist since observations demand metastability on at least $\sim$ few-year timescales for Scenario $\beta \gg 1$. Such potential mechanisms and their observational consequences, explored in \S\ref{mech}, are almost certainly predicated on self-regulation for ICDP-state systems. For the case of irradiation feedback, this is conceptually different than such feedback in LMXBs which induces mass transfer cycles; in Scenario $\beta \gg 1$ feedback on the self-excited wind stabilizes the shock until another process causes the system to transition to or from RLOF and disk states. While the irradiation flux is lower by a few orders of magnitude than in AMXPs, so is the companion mass loss rate. Irradiation feedback on the shock may also operate in SCDP-state BWs in the context of channeled particle heating rather than by photons as noted by \cite{2017arXiv170605467S} but such systems are not the focus of this work. The issue of internal companion dynamics and its influence on long-term stability is examined in \S\ref{compinternal}. Finally, we discuss potential observational discriminants of  the two scenarios in \S\ref{summary}.

\section{Scenario $\beta \ll 1$: Companion Magnetosphere Dominance}
\label{comppressure}

We consider the curved pulsar wind termination shock geometry as arising from a stellar companion magnetosphere. A strong field whose poloidal component is of order several kilogauss at the companion surface, $B_{\rm c} \gtrsim 1$~kG, is demanded for isobaric surfaces curved around the MSP in this scenario, implying a pulsar termination shock with curvature similar to the isobars. 

\subsection{General Considerations}
Throughout this work, we assume the companion is close to Roche Lobe filling and its radius $R_{\rm c}$ is approximated by the volume equivalent Roche radius $R_{\rm vL}$ of \citet{1971ARA&A...9..183P},
\begin{equation}
\frac{R_{\rm vL}}{a} = \frac{2}{3^{4/3}} \left( 1+ q \right)^{-1/3},
\label{rochepac}
\end{equation}
$R_{\rm c} \approx R_{\rm vL}$, where $q = M_{\rm p}/m_{\rm c}$ is the mass ratio.  Assuming that the companion dipolar component dominates any multipolar components at large distances from the companion, a kilogauss scale is readily derivable from a pressure balance condition for the magnetopause for an isotropic pulsar wind, e.g., \cite{1990ApJ...358..561H},
\begin{equation}
\frac{B_{\rm c}^2}{8\pi} \left(\frac{R_{\rm c}}{a - r_{\rm s}} \right)^6 = \frac{\langle S \rangle}{c}  \sim \frac{\dot{E}_{\rm SD}}{4 \pi c \, r_{\rm s}^2}
\label{Bpressurebalance1}
\end{equation}
where $r_{\rm s}$ is the characteristic shock radius as measured from the MSP, and $\langle S \rangle$ is the pulsar wind Poynting flux far outside $r_{\rm LC}$.  We define $B_{\rm 0}$ as the minimum required surface polar field $B_{\rm c}$ for $r_{\rm s}/a \leq 1/2$ from Eq.~(\ref{Bpressurebalance1}),
\begin{align}
& B_{\rm 0} \equiv \frac{a^2}{2 R_{\rm c}^3} \sqrt{ \frac{\dot{E}_{\rm SD}}{2 c}} \label{Bpmin} \\
&\approx 6 \times 10^2  \left(\frac{\dot{E}_{\rm SD}}{10^{35} \, \rm erg \, s^{-1}} \right)^{1/2} \left(\frac{P_b}{2 \times 10^4 \, \rm s} \right)^{-2/3} \nonumber  \\
& \times \left(\frac{M_{\rm p}}{1.7 \, \rm M_\odot} \right)^{-1/3} \left( \frac{1 + (3q/2)}{11.5} \right) \, \, \rm G, \quad q \gg 1 \nonumber.
\end{align}
As we demonstrate in \S\ref{isoPwind}--\ref{anisoPwind}, the companion magnetosphere scenario calls for $b \equiv B_{\rm c}/B_{\rm 0} \gtrsim 10$ (i.e. several kilogauss surface fields) for {\it{isobaric}} surfaces that are appreciably curved around the MSP in the plane of the orbit. Note the scalings of Eq.~(\ref{Bpmin}) with $q$ and $P_{b}$, which necessitate larger surface fields for lower-mass companions or shorter orbital periods. 
The isobaric surfaces are not only relevant for RBs with ICDP light curves, but also possibly for BWs and MSPs with synchronous but small quasi-degenerate companions with magnetic fields when $b <1$.

There is evidence for large, perhaps localized, kilogauss surface fields in some M dwarfs \citep{1985ApJ...299L..47S,2009ApJ...692..538R,2012LRSP....9....1R}, brown dwarfs \citep{2001Natur.410..338B} and T Tauri stars \citep{2007ApJ...664..975J}, but observational constraints of RB stellar companion fields are almost non-existent. The theoretical basis for large enduring poloidal fields is also undetermined in RB companions -- convective dynamos are poorly understood even in the Sun, and in contrast to isolated M dwarfs of similar mass, RB companions are anisotropically irradiated, highly evolved, bloated and optically brighter. Yet, the synchronous orbital rotation is faster than axial rotation in isolated M dwarfs, therefore it may be plausible for strong large-scale (rather than localized) fields to arise. Indeed, if the convective dynamo ultimately extracts its energy from the orbit tidally \citep[i.e.,][]{1992ApJ...385..621A}, then a $1$ kG poloidal field may be tidally replenished on timescales $\gg 10^6$ s without producing a large orbital period derivative violating observations. Note that the geometry of the putative poloidal component -- whether it is aligned or skew with respect to the plane formed by the orbital momentum vector and line joining the two stars -- is unknown. Due to this uncertainty, we explore arbitrary orientations.

From first principles, there are strict upper bounds on the putative poloidal field component which we treat as dipolar for practicality. Firstly, since the RB companions are tidally-locked, their dipolar fields are rotating with respect to the system barycenter. Analogous to pulsar spin-down, a repercussion is ``orbital dipole radiation" which imparts a secular torque on the companion orbit. A precise estimate is rather involved even in the vacuum limit \citep[e.g.,][]{2016MNRAS.463.1240P} and in full generality also depends on the orientation of the dipole with respect to the orbital axis. 

Note that the companion light cylinder is much larger than the binary separation, $c P_b/(2 \pi) \sim 10^{14}$ cm $\gg a$, so the system may be regarded as in the near zone, where a dipole field structure is a good approximation, for the present motivation of termination shock curvature. Moreover, orbital sweepback of companion magnetospheric field lines in the $\beta \ll1$ limit may be neglected. For a simpler order-of-magnitude estimate, we invoke the Larmor formula $\dot{T} = -\Omega_b^4 R_{\rm c}^6 B_{\rm c}^2/(6 c^3)$ where $\Omega_b$ is the orbital angular frequency; this is generally accurate within a factor of a few in comparison with force-free and dissipative MHD models \citep{2006ApJ...648L..51S,2012ApJ...754L...1K,2014ApJ...793...97K}. Then, the energy loss rate can be shown to be 
\begin{equation}
\dot{T} = -\frac{32}{3^9} \left( \frac{G M_{\rm p}}{c^2} \right)^2 \frac{B_{\rm c}^2 c}{q^2}.
\end{equation}
The concomitant characteristic timescale for orbital evolution is $\tau_{\rm mag} \sim T/|\dot{T}| \approx 300 q c^3/(G a B_{\rm c}^2)$ where $T = G M_{\rm p}^2/(2 a q)$ is the orbital kinetic energy. This rate may be compared to measured $|\dot{P}_b |/P_b \sim 10^{-15}$ s$^{-1} \gg 1/\tau_{\rm mag}$ which is attributed to the \cite{1992ApJ...385..621A} mechanism (which incidentally also posits a convective dynamo for the strong companion field; also see text following Eq.~(\ref{PbPbdot})) to yield the upper limit for $B_{\rm c}$,
\begin{equation}
B_{\rm c} \ll 10^8 \left(\frac{q}{7} \right)^{1/2} \left( \frac{a}{10^{11} \, \rm cm} \right)^{-1/2} \left( \frac{\tau_{\rm mag}}{10^{15} \, \rm s} \right)^{-1/2} \, \rm G.
\label{magbrake}
\end{equation}
In the vacuum limit, assuming there is no wind from the companion, there is an induced electric field $E_{\rm ind} \sim |\boldsymbol{v}_{\rm orb} \boldsymbol{\times} \boldsymbol{B}_{\rm c} |/c$ on the companion surface with $v_{\rm orb} = 2 \pi a q/[P_b (1+q)]$,
\begin{eqnarray}
E_{\rm ind} &\sim& 1.4 \left(\frac{q}{1+q} \right)^{2/3}  \left(\frac{B_{\rm c}}{10^3 \, \rm G} \right)  \left(\frac{M_{\rm p}}{1.7 \, \rm M_\odot} \right)^{1/3} \nonumber \\
&& \times \left(\frac{P_b}{2 \times 10^4 \, \rm s} \right)^{-1/3} \, \rm G.
\end{eqnarray}
 Analogous to the canonical pulsar case \citep[e.g.,][]{1969ApJ...157..869G}, this electrostatic force on ionized hydrogen greatly exceeds the gravitational force $F_g \sim G m_p m_{\rm c}/R_{\rm c}^2$ where $m_p$ is the mass of the proton,
 \begin{equation}
 \frac{q_e E_{\rm ind}}{F_g} \sim 10^{10}  \left(\frac{B_{\rm c}}{10^3 \, \rm G} \right)  \left(\frac{P_b}{2 \times 10^4 \, \rm s} \right) \frac{q}{(1+q)^{2/3}}.
 \end{equation}
Therefore the companion magnetosphere is plasma loaded, possibly with active currents corresponding to a global force-free MHD equilibrium. Moreover, complex current systems will also arise since pulsar wind Poynting flux distorts the companion magnetosphere, analogous to the solar wind distorting the Earth bow shock and magnetosphere. In this scenario, transitions between different force-free equilibria would manifest as bursts similar to those in the Sun \citep[e.g.,][]{2006A&A...451..319R,2008A&A...488L..71T} and also invoked for magnetar bursts \citep{2002ApJ...574..332T}, with flare emission powered by reconnection and topological changes of currents and fields.

Irradiation of the companion by the pulsar $\gamma$-rays and shock emission induces mass loss and also fills the magnetosphere with plasma, which is contained by the magnetosphere until $\beta \sim 1$. This containment timescale must be at least as long as the pulsar persistence time $\tau_{\rm p}$. Assuming the companion mass loss is similar to that inferred in the AMXP/disk states of $|\dot{m}_{\rm c}| \sim 10^{15}$ g s$^{-1}$ and that the magnetospheric reservoir  $B_{\rm c}^2 R_{\rm c}^3$ is filled at a rate $|\dot{m}_{\rm c}| v_{\rm esc}^2$ where $v_{\rm esc}$ is the isolated-star escape speed, we find a lower limit for $B_{\rm c}$,
\begin{align}
B_{\rm c} \, \gtrsim \, \, & 2 \times 10^3 \,  \left(\frac{P_b}{2 \times 10^4 \, \rm s} \right)^{-4/3}  \left(\frac{q}{7} \right)^{1/6}  \left(\frac{\tau_{\rm p}}{10^8 \, \rm s} \right)^{1/2}  \nonumber \\
\times &  \left(\frac{M_{\rm p}}{1.7 \, \rm M_\odot} \right)^{-1/6}    \left(\frac{|\dot{m}_{\rm c}|}{10^{15} \rm \, g \, s^{-1} } \right)^{1/2} \rm \quad G.
\label{Breservoir1}
\end{align}
Similarly, if mass loss exists via RLOF rather than a wind, then for ion thermal speed $c_s \sim 10^6$ cm s$^{-1}$ and a higher rate $|\dot{m}_{\rm c}| \sim 10^{16.5}$ g s$^{-1}$,
\begin{align}
B_{\rm c} \, \gtrsim \, \, & 3 \times 10^2 \,  \left(\frac{P_b}{2 \times 10^4 \, \rm s} \right)^{-1} \left(\frac{\tau_{\rm p}}{10^8 \, \rm s} \right)^{1/2}  \left(\frac{M_{\rm p}}{1.7 \, \rm M_\odot} \right)^{-1/2} \nonumber \\
\times &  \left(\frac{q}{7} \right)^{1/2}      \left(\frac{|\dot{m}_{\rm c}|}{10^{16.5} \rm \, g \, s^{-1} } \right)^{1/2}  \left(\frac{c_s}{10^{6} \rm \, cm \, s^{-1} } \right)  \rm \quad G.
\label{Breservoir2}
\end{align}
Therefore, strong fields are also essential if there is significant mass loss during the pulsar state (see \S\ref{masscons} for constraints) for magnetic dominance to be sustained. 

A similar-in-magnitude constraint on $B_{\rm c}$ to Eq.~(\ref{magbrake}) may be established by noting that the energy of the putative poloidal field must be a small fraction of the gravitational binding energy, $B_{\rm c}^2 R_{\rm c}^3 \ll G m_{\rm c}^2/R_{\rm c}$. This implies
\begin{equation}
B_{\rm c} \ll 10^8 \left(\frac{P_b}{2 \times 10^4 \, \rm s} \right)^{-4/3}  \left(\frac{q}{7} \right)^{-1/3}  \left(\frac{M_{\rm p}}{1.7 \, \rm M_\odot} \right)^{1/3}  \, \rm G.
\label{bindingerg}
\end{equation}
Another restriction is that gas dominance, $\beta_{\rm c}  = 8 \pi n k_b T_{\rm c}/B_{\rm c}^2 \gg 1$, is required inside the star for the convective dynamo to exist. Estimating the density as the average value, $\langle n \rangle m_p \approx 3^5 \pi/(8 G P_b^2)$ for a pure hydrogen atmosphere, the temperature at the photosphere and the magnetic field as the surface value, we find,
\begin{align}
B_{\rm c} \ll & \, \, 7 \times 10^6  \left(\frac{T_{\rm c}}{6000 \, \rm K} \right)^{1/2} \left(\frac{P_b}{2 \times 10^4 \, \rm s} \right)^{-1} \nonumber \\
&\times \left(\frac{\beta_{\rm c}}{1} \right)^{-1/2} \, \rm G.
\label{plasmabeta1}
\end{align}
Areas of the photosphere may attain $\beta_{\rm c} \sim 1$, as in the active Sun and magnetically active stars. At the photosphere, the number density is approximately $n \sim \tau/(m_p \kappa H)$ where $\tau$ is the optical depth, $H = k_b T_{\rm c}/(m_p g_c)$ is the pressure scale height, $\kappa$ the continuum opacity and $g_c = G m_{\rm c}/R_{\rm c}^2$ the isolated-star surface gravity. In Thomson electron scattering limit, dominant for very high plasma temperatures \citep{1983psen.book.....C}, we have $\kappa \rightarrow \kappa_{\rm es} \approx 0.4$ cm$^{2}$ g$^{-1}$ and $B_{\rm c}$ is limited to kilogauss fields when $\beta_{\rm c} \gtrsim 1$ and $\tau$ is moderately large,
\begin{align}
B_{\rm c}^{\rm es} \lesssim & \, \, 1.4 \times 10^3 \,  \left(\frac{P_b}{2 \times 10^4 \, \rm s} \right)^{-2/3}  \left(\frac{M_{\rm p}}{1.7 \, \rm M_\odot} \right)^{1/6}  \left(\frac{q}{7} \right)^{-1/6} \nonumber \\
&  \times \left(\frac{\beta_{\rm c}}{1} \right)^{-1/2}  \left(\frac{\tau}{1} \right)^{1/2}  \quad \rm G.
\label{plasmabetaes}
\end{align}
If the mean opacity follows a Kramer law, $\kappa_{\rm K} = \kappa_{\rm K}^0 m_p n T^{-7/2}$, then
\begin{align}
B_{\rm c}^{\rm K} \lesssim & \, \, 2 \times 10^2 \left( \frac{\kappa_{\rm K}^0 }{10^{23} \, \, \rm cm^2 \, g^{-1} } \right)^{-1/4}  \left(\frac{T_{\rm c}}{6000 \, \rm K} \right)^{9/8}   \nonumber \\
& \times  \left(\frac{P_b}{2 \times 10^4 \, \rm s} \right)^{-1/3} \left(\frac{M_{\rm p}}{1.7 \, \rm M_\odot} \right)^{1/12}   \left(\frac{q}{7} \right)^{-1/12}  \nonumber \\
 & \times  \left(\frac{\beta_{\rm c}}{1} \right)^{-1/2}  \left(\frac{\tau}{1} \right)^{1/4} \quad \rm G.
\label{plasmabetaK}
\end{align}
For a value of $\kappa_{\rm K}^0 \approx 10^{23} \, \, \rm cm^2 \, g^{-1}$ typical of bound and free-free absorption \citep{1983psen.book.....C}, apparently $B_{\rm c}$ necessary for pressure balance are generally excluded. That is, moderately hot companions with $T \sim \rm few \times 10^4$ K where a Kramer law dominates ought not to be found in ICDP RBs systems in the magnetospheric scenario for pressure balance. Pressure balance in the magnetospheric scenario also advocates for much lower opacities (at temperatures below where a Kramer law operates) due to $H^{-}$ bound-free transitions and metals \citep{2005oasp.book.....G} to be the dominant sources of continuum opacity for the temperature range of interest in RB atmospheres. We also note that for the quiet Sun, $\beta_\odot \sim 10^2$--$10^4$ in the photosphere and chromosphere \citep{2017ApJ...850L..29B}; if similar values are realized in RB companions, then $B_{\rm c}$ cannot attain the kilogauss fields necessary for pressure balance.

 The constraints Eqs.~(\ref{magbrake}), (\ref{Breservoir1})--(\ref{plasmabeta1}) along with Eq.~(\ref{Bpmin}) for pressure balance form an allowed region in the $P_b, q$ and $\dot{E}_{\rm SD}$ parameter space for this scenario's tenability. Far from the shock in the companion magnetosphere, $\beta \ll 1$ for ions is fulfilled for any reasonable plasma number density provided that the ion temperature $\ll 10^9$ K. This is readily found from pressure balance Eq.~(\ref{Bpressurebalance1}), where the magnetic field at the shock scales as $B_{\rm s} \sim \sqrt{2\dot{E}_{\rm SD}/c}/r_{\rm s}$, prior to any modification by MHD jump conditions \citep[e.g.,][]{1984ApJ...283..694K}. Finally, we note that the electromagnetic forces on ions dominate any gravitational influences from the pulsar, i.e. $G m_p M_{\rm p}/r_{\rm s}^2 \ll q_e (c_s/c) B_{\rm s}$ is satisfied provided that,
\begin{align}
r_{\rm s} \gg &  \frac{G M_{\rm p} m_{\rm p} c^{3/2}}{\sqrt{2 \dot{E}_{\rm SD}} q_e c_{s}} \, \sim  \,  10^5 \, \left(\frac{\dot{E}_{\rm SD}}{10^{35} \, \rm erg \, s^{-1}} \right)^{-1/2}  \\
\times & \left(\frac{M_{\rm p}}{1.7 \, \rm M_\odot} \right)  \left(\frac{c_s}{10^{6} \, \rm cm \, s^{-1}} \right)^{-1}   \left(\frac{\sigma}{10^{-2}} \right)^{-1/2}   \, \rm cm. \nonumber
\label{rsggmag}
\end{align}
which is much smaller than $a$. Therefore, gravitational influences of the MSP are negligible on the local plasma dynamics unless the ions are unjustifiably cold. Likewise, it can be shown that Coriolis influences are negligible on the local plasma dynamics. Gravitational influences could become consequential on the macroscopic (i.e. fluid description) plasma dynamics, but it depends on the details of the MHD equilibria and currents induced in the magnetosphere. 

\begin{figure*}[th]
\centering
\includegraphics[width=0.85\textwidth]{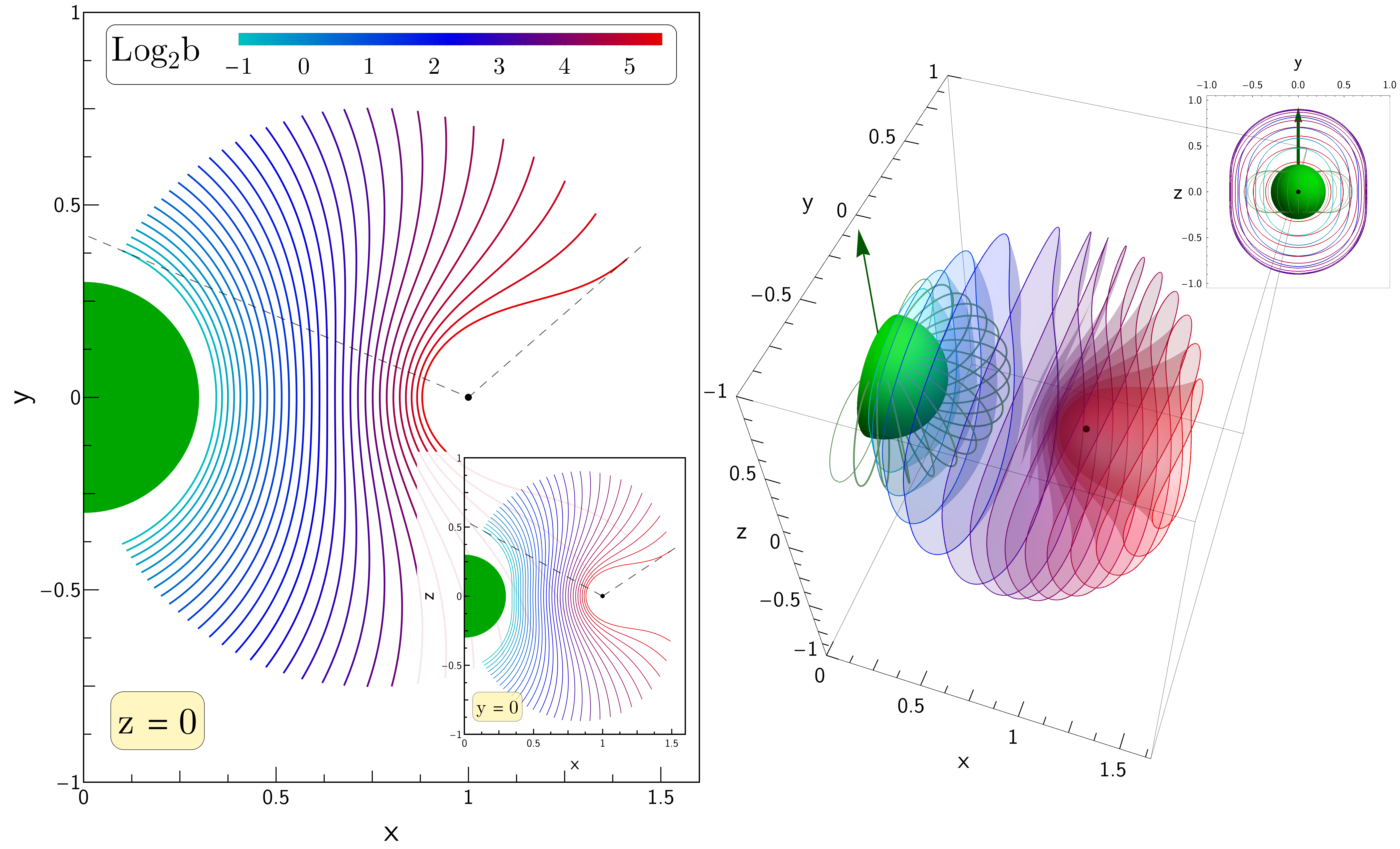}
\caption{Isobaric surfaces, to scale, for a companion star dipolar magnetosphere aligned along the $\boldsymbol{\hat{z}}$ axis, defined by Eq.~(\ref{singleshockcond})-(\ref{Bzdipole}). The green disk represents the companion of typical radius $R_c \approx 0.3a$ at the origin, while the black dot is the MSP at $\{1,0,0\}$. For $b =  B_{\rm c}/B_{\rm 0} \gtrsim 2^3$, the isobaric surface begins to appreciably curve around the MSP. {\bf{Left:}} Cut in $x-y$ and $x-z$ planes of the surface, demonstrating the curved nature of the isobaric surface for color-coded $b$ as indicated by the scale. The dashed black line highlights the boundary condition imposed by Eq.~(\ref{singleshockcond}), and establishes that the surface boundary is indeed tangential to the radial pulsar outflow. {\bf{Right:}} A 3D representation of the surface, with inset of the $y-z$ projection highlighting this symmetrical case. }
\label{Bzdipolefig}
\end{figure*}

\subsection{Isobaric Surfaces}
\label{surfaces}

Beyond the generic considerations above, we now explore pressure balance of a 3D dipolar companion field by a relativistic magnetized pulsar wind, analytically described as a Poynting flux. On radial length scales much larger than $r_{\rm LC}$, or time-averaged over timescales much longer than the pulsar period, the pulsar wind is asymptotically radially outflowing and toroidal-field dominated. A simpler isotropic case considered in \S\ref{isoPwind} preludes to the more complex anisotropic case in \S\ref{anisoPwind}.

There are several caveats to our rudimentary considerations of pressure confinement of the pulsar wind or companion magnetosphere. The isobaric surface geometry describes where the extended shock structure ought to exist, but does not appraise any backreactions on the companion magnetosphere or pulsar wind. For instance, the companion magnetosphere will be severely distorted away from the vacuum dipole towards distinct non-potential force-free MHD equilibria. The gas pressure may also be dynamically important near the companion surface, i.e. where $\beta_{\rm c} \gtrsim 1$. Moreover, the termination shock itself is a significant region of conversion of the magnetic wind into particle energy, which will influence the global shock structures. Even in the hydrodynamic limit, there is some thickness to the overall shock structure: the pulsar termination shock, followed by an astropause and contact discontinuity \citep[e.g.,][]{2016A&A...586A.111S}. In an MHD formalism with two or more fluid species (e.g., a pair plasma interacting with an electron-ion plasma), many different wave modes may be excited leading to complex interposing shock structures \citep{2010adma.book.....G}. Such rich complexity is exhibited in relativistic MHD simulations of pulsar winds \citep[e.g.,][]{2005A&A...434..189B,2018arXiv180407327B}, yet pressure balance/confinement of the pulsar wind remains a credible estimate of the global structure, particularly proximate to the shock apex. The termination shock head, where the particle acceleration occurs, is largely what is relevant for ICDP light curves in RBs \citep{2017ApJ...839...80W}, rather than peripheral regions of the shock structures. Therefore, we focus on such pressure surfaces and defer global MHD simulations to future studies.

Aside from pressure balance, the orientation of the companion dipolar magnetosphere will also influence the efficiency and locales of relativistic particle acceleration in the termination shock. How is rather unclear and is deferred to future kinetic studies. Large-scale dipolar fields may also introduce peculiar orbital phase-dependent polarization character on the synchrotron ICDP light curves. 

\subsubsection{ Isotropic Pulsar Wind}
\label{isoPwind}

The companion dipolar field component in spherical polar coordinates with origin at the companion center is given by
\begin{equation}
\boldsymbol{B} = \frac{B_{\rm c}}{2} \left( \frac{ R_{\rm c}}{a \, r} \right)^3 \left( 2 \cos \theta \, \boldsymbol{\hat{r}} + \sin \theta \, \boldsymbol{\hat{\theta}} \right)
\end{equation}
with $r$ dimensionless in units of $a$ and $\theta$ is a polar angle. Due to the lack of symmetry of the pressure balance condition for an arbitrarily oriented dipole, it is more convenient to work in Cartesian coordinates normalized to units of $a$. We define the $\boldsymbol{\hat{z}}$ and $\boldsymbol{\hat{x}}$ as parallel with the orbital momentum vector and line joining the two stars, respectively, with the pulsar at $x = 1$ and companion at the origin. The implicit isobaric surface of the companion magnetosphere and pulsar wind Poynting flux is given by the 3D generalization of Eq~(\ref{Bpressurebalance1}) with scalar field of pressure $\cal{G}$,
\begin{equation}
0= {\cal{G}}(x,y,z) = \frac{\left |\boldsymbol{B}^2 \right|}{8 \pi} -\frac{\langle S \rangle}{c} 
\label{calG}
\end{equation}
where for an isotropic pulsar wind
\begin{equation}
\left(\frac{\langle S \rangle}{c}\right)_{\rm iso} =  \frac{\dot{E}_{\rm SD}}{4 \pi c \, a^2 |\boldsymbol{r_{\rm p}} |^2}
\end{equation}
and $\boldsymbol{r_{\rm p}} \equiv \{x-1,y,z\}$ is the outward radial vector from the pulsar. For simplicity, we also impose the condition,
\begin{equation}
\boldsymbol{\nabla} {\cal G} \, \boldsymbol{ \cdot \, r_{\rm p}} > 0
\label{singleshockcond}
\end{equation}
that precludes multivaluedness of the pressure surface, i.e. physically, there is a single termination shock for the putatively radial pulsar wind. Beyond the boundary imposed by this condition, the interaction geometry is indeterminate but with the radial pulsar Poynting flux dominating far from the boundary locale.

\begin{figure}[t]
\centering
\includegraphics[width=0.45\textwidth]{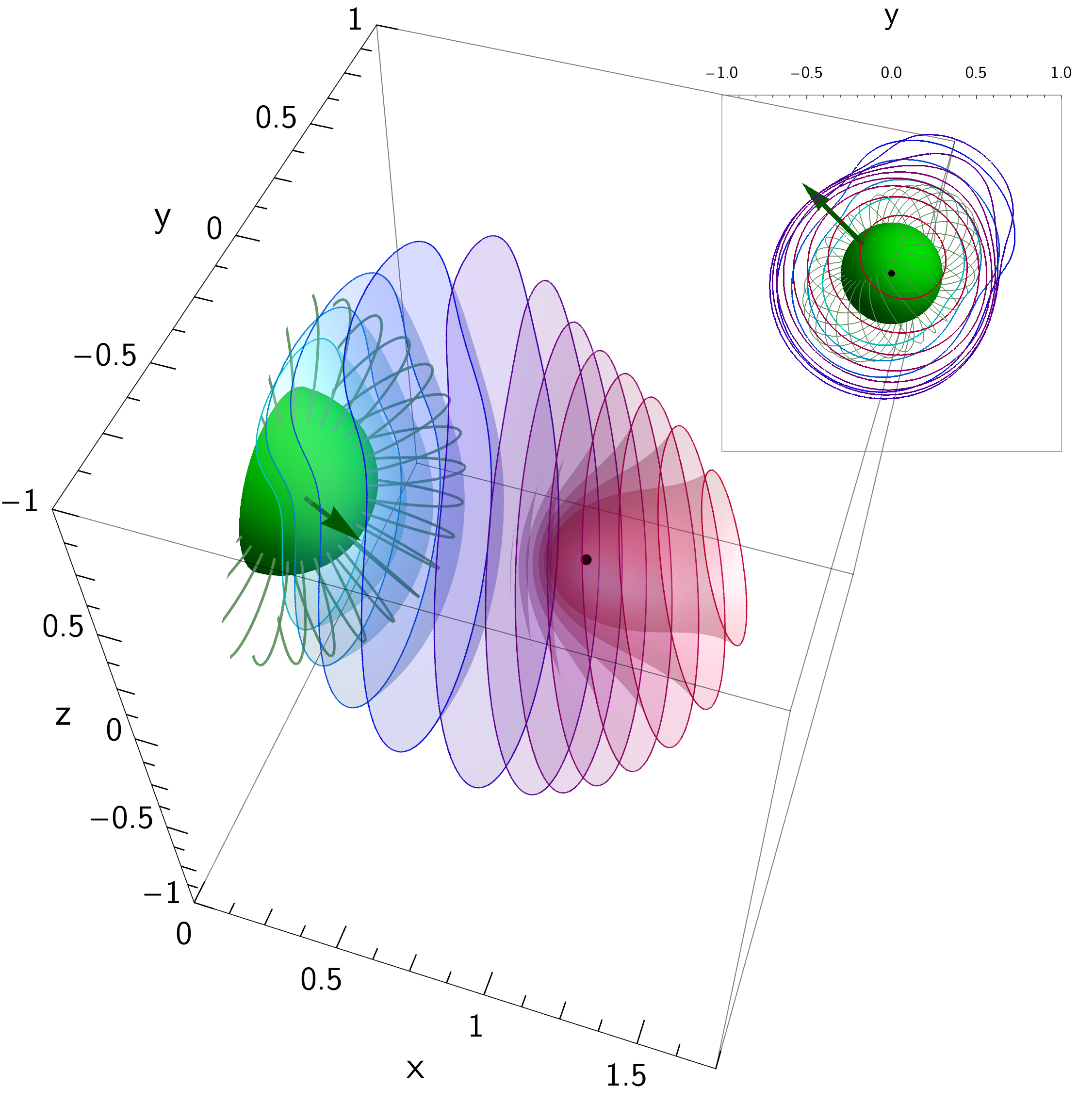}
\caption{3D isobaric surfaces for a companion star dipolar magnetosphere whose dipole moment vector axis coincides with the green arrow; undistorted field lines for a vacuum dipole are also depicted. The color coding of $b$ is identical to that of  Fig.~\ref{Bzdipolefig}. In contrast to Fig.~\ref{Bzdipolefig}, there are clear  asymmetries in the surface geometry. Nevertheless, close to the stagnation point, the geometry for large $b$ is approximately hemispherical. Inset: orthographic projection on the $y-z$ plane.}
\label{arbdipolefig}
\end{figure}

For concreteness, consider a dipole whose axis is $\boldsymbol{\hat{z}}$. Then, the isobaric surface of the companion magnetosphere and MSP Poynting flux may be shown to be implicitly defined by
\begin{align}
 {\cal{G}}_z =& \, \,  b^2 \left[(x-1)^2 +y^2+z^2 \right] (x^2+y^2+4 z^2) \nonumber \\
& - 64 (x^2+y^2+z^2)^4 \stackrel{!}{=} 0
\label{Bzdipole}
\end{align}
where $b \equiv B_{\rm c}/B_{\rm 0}$. We plot Eq.~(\ref{Bzdipole}) in Figure~\ref{Bzdipolefig} with color-coded values of $b$. For smaller values of $b \lesssim 1$, the isobars are curved around the companion as expected. For larger values of $b \gtrsim 2^3$, there is clear curvature of isobars around the MSP particularly near the magnetopause; this depiction is analogous to Fig~1 in \cite{2017ApJ...839...80W}. These larger values of $b \gtrsim 2^3$ are required for significant curvature of the shock head, particularly in the $z=0$ plane. Such geometric curvature is central to the X-ray orbital modulation observed in RBs, and in models of such emission the observed double-peak phase separation couples to the putative shock opening angle \citep{2017ApJ...839...80W}. Even larger $b$ are not shown, as they yield a total envelopment of the pulsar and may violate Eq.~(\ref{plasmabeta1}). Such envelopment, however, could lead to prolific reconnection events and flares behind the pulsar (i.e. $x>1)$. Some X-ray flares \citep{2018arXiv180900215C} and mini radio eclipses of the MSP at pulsar inferior conjunction are observed in a some RBs \citep[e.g.,][]{2015ApJ...800L..12R}, but not contemporaneously in the same system.

Note that in the peculiar case of Eq.~(\ref{Bzdipole}), there is symmetry about the $y$ and $z$ axes. For a dipole with axis along $\boldsymbol{\hat{x}}$, the surfaces (not shown) even exhibit azimuthal symmetry about $\boldsymbol{\hat{x}}$. In general, there are no such symmetries for the isobaric surface for an arbitrarily oriented dipolar field for even an isotropic pulsar wind. One such skewed-dipole illustrative case is depicted in Figure~\ref{arbdipolefig}. Close to the stagnation point when $b\gtrsim 2^4$, the head of the shock region is approximately hemispherical, a consequence of the isotropic pulsar wind considered in this section. However,  at locales of the boundaries defined by Eq.~(\ref{singleshockcond}), there are clear asymmetries. Such asymmetries may account for the small phase offset from IC in some ICDP systems as well as apparent asymmetries about IC in ICDP light curves. This forms an alternative scenario to Coriolis effects of a companion wind invoked in the past \citep{2016arXiv160603518R,2017ApJ...839...80W} and for Scenario $\beta \gg 1$.

\begin{figure*}[th]
\centering
\includegraphics[width=0.975\textwidth]{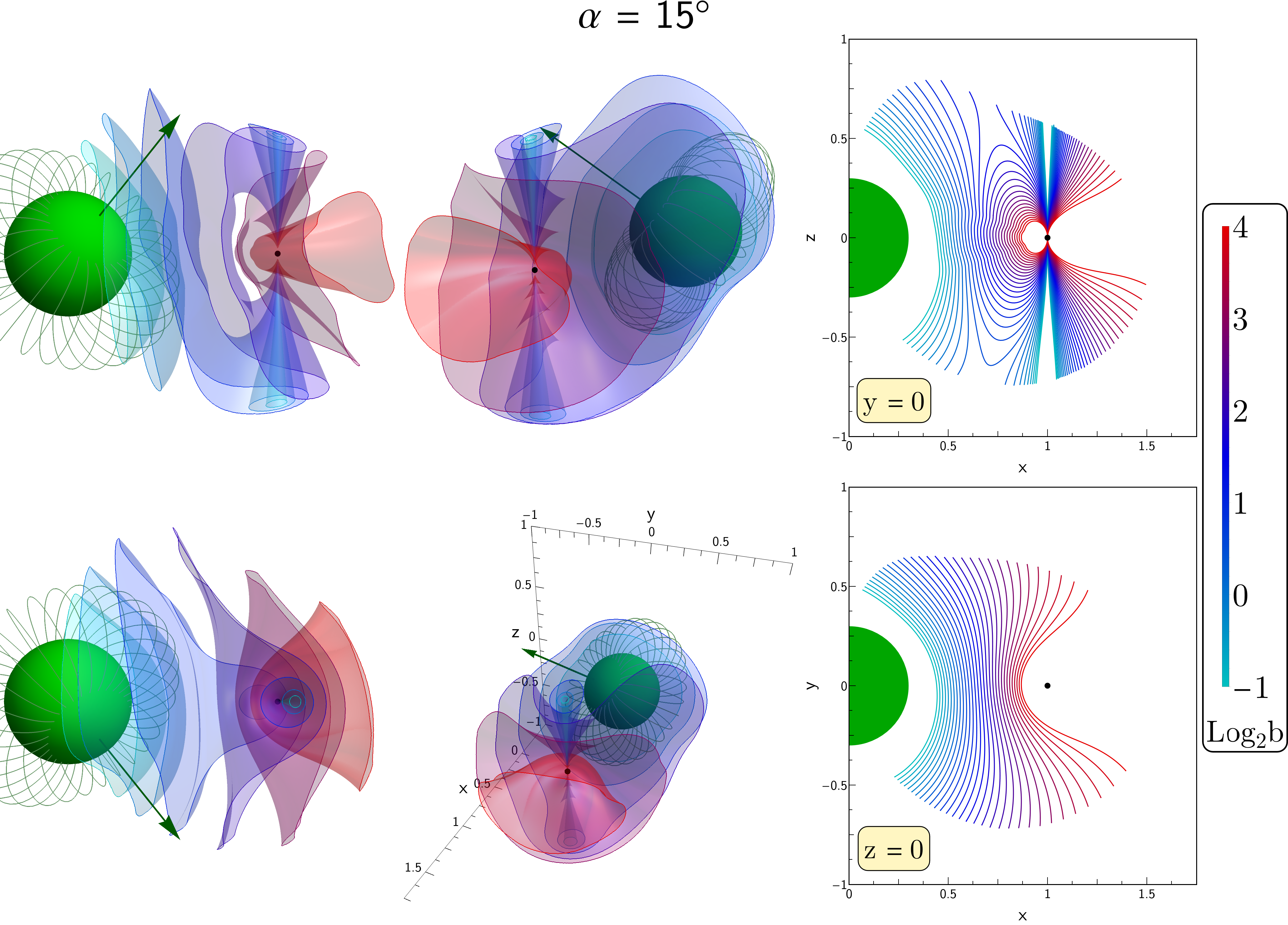}
\caption{Views of isobaric surfaces for an anisotropic pulsar wind with $\alpha =15^\circ$ and misaligned dipolar magnetosphere of the companion. As in Figs~\ref{Bzdipolefig} and \ref{arbdipolefig}, the parameter $b$ scales the companion poloidal field in units of $B_{\rm 0}$, but with a different color scaling range for clarity.  The leftmost two columns depict different views of the isobars. The rightmost column depicts cuts in the $x-y$ and $x-z$ planes through the origin. Funnel-like isobaric surfaces, likely a extraneous solution regime, exist along the pulsar spin axis. See text for details.}
\label{Aniso15angled}
\end{figure*}

\begin{figure*}[th] 
\centering
\includegraphics[width=0.975\textwidth]{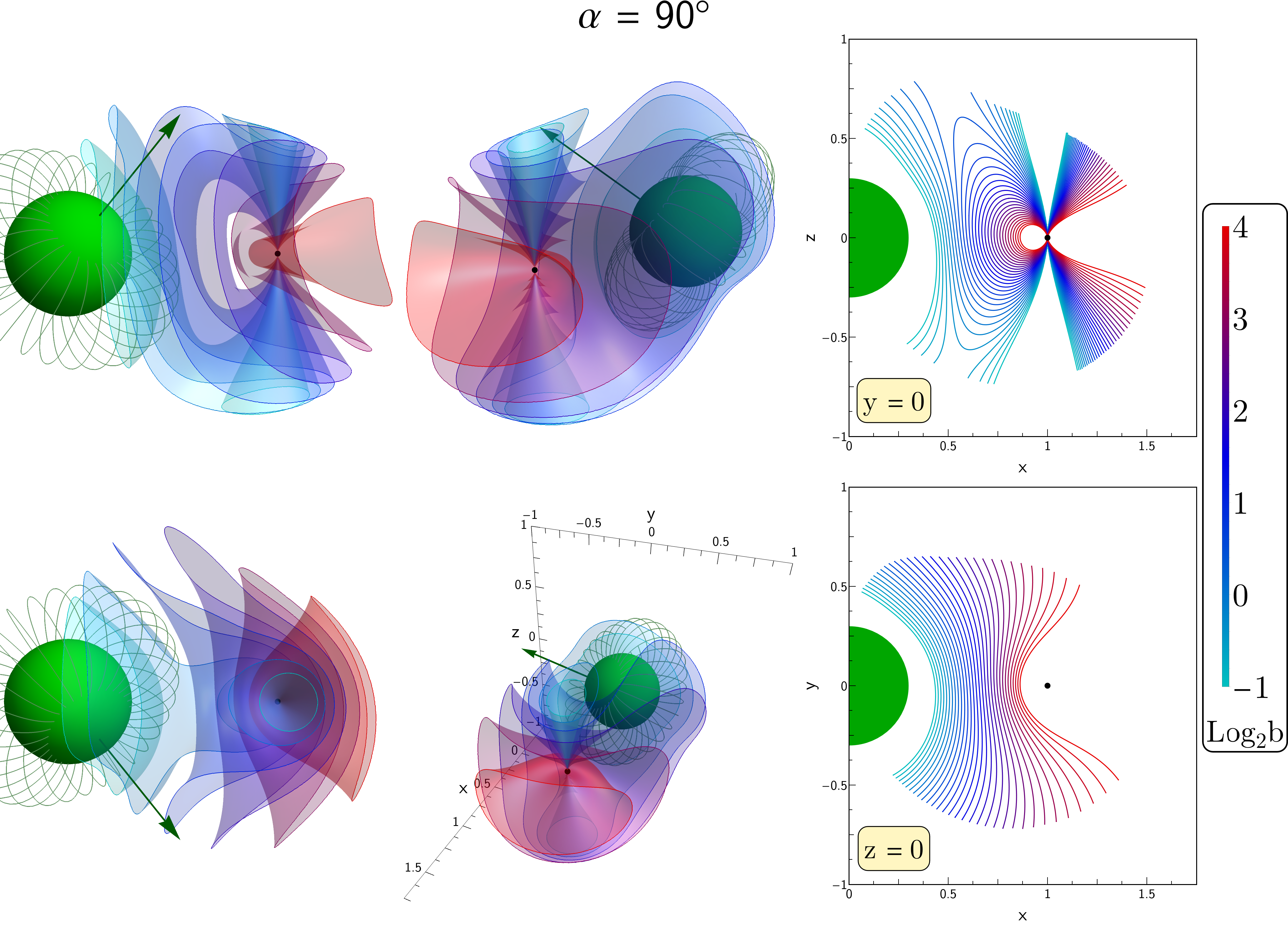}
\caption{Views of isobaric surfaces with identical construction as Figure~\ref{Aniso15angled}, but with $\alpha=90^\circ$. }
\label{Aniso90angled}
\end{figure*}

\subsubsection{Anisotropic Pulsar Wind}
\label{anisoPwind}

Soon after the discovery of pulsars, the Poynting flux of pulsar winds were widely recognized to be anisotropic \citep{1969ApJ...158..727M,1973ApJ...180..207M} and plasma-loaded \citep{1969ApJ...157..869G}. In the force-free limit of a plasma-filled magnetosphere, the anisotropy of the pulsar wind Poynting flux is contingent on the magnetic obliquity $\alpha$ of the rotator and roughly varies between $\sin^2 \vartheta$ (aligned rotator) and to $\sin^4 \vartheta$ (orthogonal rotator) \citep{1999A&A...349.1017B}. Here, $\vartheta$ is the polar angle with respect to the spin axis, i.e., $\cos \vartheta = z/|\boldsymbol{r_{\rm p}}|$. Moreover, for the orbital scales of interest in BWs and RBs, the azimuthal anisotropies on the scale of $r_{\rm LC}$ may be neglected (i.e. we restrict to the far zone). 
For simplicity, we consider the pulsar spin axis aligned with orbital axis $\hat{z}$. Such alignment is expected from the formation/evolution recycling scenario for MSPs, as hinted for RB J2215+5135 and other MSPs \citep{2014MNRAS.439.2033G,2014ApJS..213....6J}, and as known for other stellar contexts \citep{2007A&A...474..565A,2011MNRAS.413L..71W}. 

Using a sample of force-free MHD simulations which are appropriate for the gross global structure of the pulsar wind, \cite{2016MNRAS.457.3384T} analytically parameterized the $\alpha$ dependence of asymptotic magnetized pulsar winds far outside $r_{\rm LC}$ as a sum of the wind structure of aligned and orthogonal rotators. Their semi-analytic construction is accurate to within $\sim 10\%$ to simulations for the differential Poynting flux averaged over azimuthal angles. From their prescription, we obtain a convenient expression of the azimuthally-averaged anisotropic differential (in solid angle) Poynting flux in terms of the observable $\dot{E}_{\rm SD}$  after a modicum of algebra,  
\begin{equation}
\left(\frac{\langle S \rangle}{c}\right)_{\rm aniso} = \frac{2\dot{E}_{\rm SD}}{4 \pi c \, a^2 |\boldsymbol{r_{\rm p}} |^2} \frac{\langle {\cal{B}}^2 \rangle_\phi}{C_0} \sin^2 \vartheta 
\label{anisowindanalytic}
\end{equation}
where
\begin{equation}
\langle {\cal{B}}^2 \rangle_\phi = \frac{1}{2 \pi} \int_0^{2 \pi}  {\cal{B} (\alpha, \phi)}^2 \, d \phi
\end{equation}
and 
\begin{align}
 {\cal{B}} \approx & \, \,  {\cal{B}}_{\parallel} + {\cal{B}}_{\perp}  \label{calB} \\
  {\cal{B}}_{\parallel} = &  \left[1 + 0.02 \sin \gamma + 0.22(|\cos \gamma| - 1) \right. \nonumber \\
& \left.- 0.07 (| \cos \gamma | - 1)^4 \right]  | 1 - 2 \alpha/\pi | \sign(\cos \gamma)  \nonumber \\
   {\cal{B}}_{\perp} = & \left(1 + 0.17 \left| \sin 2 \alpha \right| -  | 1 - 2 \alpha/\pi | \right) \nonumber \\
  & \times \sin \vartheta \cos \left( \phi - \frac{\pi}{6} \right) \nonumber.
\end{align}
Here, ${\cal{B}}$ is related to the radial magnetic field $B_{r}$ in \cite{2016MNRAS.457.3384T}, with $\langle {\cal{B}}^2 \rangle_\phi$ the azimuthal angle average, and $\gamma$ the magnetic colatitude, $ \cos \gamma = \sin \alpha \sin \vartheta \cos \phi + \cos \vartheta \cos \alpha$. 
The constant $C_0$ is of order unity and normalizes the total Poynting flux integrated over solid angles,
\begin{equation}
C_0 = \int_0^{\pi}  \langle {\cal{B}}^2 \rangle_\phi \sin^3 \vartheta \, d \vartheta .
\end{equation}
Numerically, $C_0 \approx \{0.834, 0.898, 1.03, 0.768 \}$ when $\alpha = \{15^\circ, 30^\circ, 60^\circ, 90^\circ \}$, respectively. 

The form of Eq.~(\ref{anisowindanalytic}) allows for analogous nondimensionalization of the pressure balance condition Eq.~(\ref{calG})  as Eq.~(\ref{Bzdipole}) after some algebra. The condition Eq.~(\ref{singleshockcond}) is more involved because of spatial derivatives of numerical integrals and is computed semi-analytically. Then, computation of implicit isobaric surfaces follows routinely. We do not consider the $\alpha = 0$ case, as the neutron stars in RBs are pulsars.

Figure~\ref{Aniso15angled} depicts computed isobaric surfaces for $\alpha =15^\circ$, comparable to the low magnetic obliquity inferred for RB J2215+5135 \citep{2014ApJS..213....6J}, and with an arbitrarily skewed companion dipole moment.  There are several intriguing features worth highlighting, in comparison to the isotropic pulsar wind cases explored in \S\ref{isoPwind}. Firstly, for low values of $b$, the isobaric surfaces are largely similar in form, since anisotropies are less pronounced in the plane of the orbit where $\sin^2 \vartheta \sim 1$. For larger values of $b \gtrsim 1$, there is a dramatic shift in the topology of the surfaces principally due to the $\sin^2 \vartheta$ factor in Eq.~(\ref{anisowindanalytic}) which guarantees a region of very low wind pressure along the spin axis $\boldsymbol{\hat{z}}$. This leads to the pronounced ``spin axis funnels", some of which are disjointed from the shock head and tail in the $y=0$ plane owing to the condition Eq.~(\ref{singleshockcond}). Moreover, for a skewed companion dipole moment, there are regimes of moderate $b\sim 2-3$ where the spin axis funnels are only partially disjointed. Clearly, regimes may be also realized where $b$ is critical between a joined and unjoined topology. In this critical regime, magnetic reconnection and transient phenomena ought to be prolific.

In Figure~\ref{Aniso90angled}, we depict computed isobaric surfaces with skewed companion dipole moment identical to that as Figure~\ref{Aniso15angled} but with $\alpha = 90^\circ$. The topology of these surfaces is largely indistinguishable to the $\alpha = 15^\circ$ case, particularly for the shock head when $b \gtrsim 2^3$ which putatively governs the X-ray orbital modulation, implying an insensitivity of isobars with large disparities of pulsar $\alpha$. Indeed, there is negligible variance in the $z=0$ plane of the two $\alpha$ cases where $\sin \theta \sim 1$. Yet, the spin axis funnel for $\alpha = 90^\circ$ exhibits a much wider opening angle, due to the stronger $\sin^4 \vartheta$ anisotropy of the pulsar wind. This is suggestive that sporadic accretion may be easier for more orthogonal rotators.

For lower values of $b$, it may be argued that these ``spin axis funnels" are entirely spurious since they are disjointed from the principal isobaric surfaces near the companion and therefore current closure (in the force-free limit) is inhibited. Likewise, for larger values of $b$, the funnels will be disrupted by the reflected back-flowing pulsar wind from the termination shock. Relativistic MHD simulations, and possibly kinetic ones as well, are required to assess the character of the funnels and shock structures as $b$ varies. Yet, relativistic MHD simulations of anisotropic pulsar wind shock interactions \citep[e.g.,][]{2004MNRAS.349..779K,2018arXiv180407327B} indicate some reality to the funnel-like structures along the pulsar spin axis, as suggested by observations of the Crab plerion \citep{2000ApJ...536L..81W,2017hsn..book.2159S,2017JPhCS.932a2050K}. Speculatively, for larger values of $b$, the surfaces are connected implying threading of the companion magnetosphere into the funnel which may be paths for sporadic accretion onto one or both poles of the MSP initiated by transitions of different force-free field configurations of the companion. Indeed, joined funnels may play a role in recently observed enhanced spin-down torques on J1023+0038 in a AMXP state \citep{2016ApJ...830..122J} where the assumption of $\beta \ll 1$ breaks down close to the MSP.

\section{Scenario $\beta \gg 1$: Quasi-Hemispherical Gravitational Capture of Companion Mass Loss by the Pulsar}
\label{gravsec}

State transitions of some RBs to AMXP-like accretion disk states implies efficient angular momentum transport of the companion mass loss. Such disk states are evidently regimes of $\beta \gtrsim 1$ for the companion mass loss. Therefore, we are motivated to examine whether companion mass loss without dynamical influences of magnetic fields may yield a stable shock curved around the pulsar rather than a disk in the rotation-powered state. Without gravitational influences of the MSP, it is generally accepted that the companion wind overpowering the MSP wind is energetically untenable on long timescales. However, we suggest that if there is sufficient angular momentum loss of the companion wind, a shock curved around the MSP may be attainable. The stability of such a putative configuration is questionable, and we explore mechanisms that may provide stability.

Here, we assume the donor is near but not entirely Roche lobe filling in the rotational-powered state {so that high mass loss rates $| \dot{m}_{\rm c}| \sim 10^{15}-10^{16}$ g s$^{-1}$ are attainable} without a disk as in conventional RLOF. The relatively high mass loss rates required by this scenario currently do not violate any observational constraints (see \ref{masscons}).

\subsection{The Circularization Radius}
\label{circradsec}

In this Section, we show that for evaporative winds from the companion, the existence of a shock implies a lower bound on the companion mass. This is a rather general result if angular momentum loss of the companion wind occurs far from the launching point which is putatively near the companion photosphere. 

In the absence of a strong companion magnetosphere, a stipulation for a shock to exist bowed around the pulsar rather than a disk is that the wind characteristic circularization radius $r_{\rm circ}$ be small compared to the characteristic shock radius $r_{\rm s}$, for a companion wind with speed $v_{\rm w}$. The circularization radius is defined by where the specific angular momentum at the accretion radius $r_{\rm acc} = 2 G M_{\rm p}/v_{\rm w}^2$ is equal to that for a Keplerian orbit at radius $r_{\rm circ}$, i.e. $(1/4) r_{\rm acc}^2 \Omega_{\rm b} \approx \sqrt{G M_{\rm p} r_{\rm circ}}$ where $ \Omega_{\rm b}$ is the orbital angular speed of the system and $M_{\rm p}$ the MSP mass \citep{1976ApJ...204..555S,2002apa..book.....F}. Therefore, $r_{\rm circ}$ is the lengthscale within which one may expect a disk to exist. This definition exhibits a strong scaling on the wind speed $v_{\rm w}$,
\begin{equation}
\frac{r_{\rm circ}}{a} \approx \frac{1}{16}\left(\frac{r_{\rm acc}}{a} \right)^4 \left(\frac{1+q}{q} \right) \approx \left( \frac{ v_{\rm orb}}{v_{\rm w}} \right)^8  \left(\frac{1+q}{q} \right)^5 \,  ,
\label{rcirc}
\end{equation}
where $v_{\rm orb} = \sqrt{G M_{\rm p}/a} \sqrt{q/(1+q)} $ is the orbital speed of the secondary. The ratio $v_{\rm w}/v_{\rm orb}$ is the characteristic Rossby number of the secondary's wind.

Parametrizing the stellar wind as a scaling of the isolated-star escape speed, $v_{\rm w}^2 = \lambda v_{\rm esc}^2$, and casting the secondary stellar radius as a fraction ${\cal{F}} \equiv R_{\rm c}/R_{\rm vL}$ of the characteristic volumetric Roche radius from Eq.~(\ref{rochepac}), one arrives at $r_{\rm circ}/a$ being a simple function of $q$ and ratio $\mu = {\cal{F}}/\lambda$,
\begin{equation}
\frac{r_{\rm circ}}{a} \approx 3 \times 10^{-3} \, \, \,  \mu^4  \frac{q^3}{(1+q)^{1/3}}.
\label{rcirc2}
\end{equation}
The typical thermal speed is $c_{\rm s} \sim 10^6$ cm s$^{-1}$ for $T \sim 10^4$ K while an irradiation-induced evaporative wind speed $v_{\rm w} \gg c_s$ may be on the order of the escape speed of the companion $v_{\rm esc} \lesssim \sqrt{2 G m_{\rm c}/R_{\rm c}} \sim 5 \times 10^{7}$ cm s$^{-1} \gg c_{\rm s}$ for a typical RB secondary of mass $m_{\rm c} \approx 0.3 M_\odot$ and radius $R_{\rm c} \approx 0.4 R_\odot \approx 3 \times 10^{10}$ cm. This $v_{\rm esc}$ is an upper limit to $v_{\rm w}$, since for a star near the Roche limit $v_{\rm esc}$ may be substantially lower owing to the low potential barrier. Coincidentally, $v_{\rm esc}$ is also on the order of the escape speed from the entire system $ \sim \sqrt{2 G M_{\rm p}/a}$ or the orbital speed of the companion. 

Requiring $r_{\rm circ}/a \lesssim 0.5$ for $\mu =1$, since the putative shock exists past this point, then implies $q \lesssim 7$ which clearly excludes some lower-mass RB companions; therefore this calls for $\mu \lesssim 1$ or $\lambda \gtrsim 1$. This restriction is not very constraining due to the strong fourth-power dependence of $\mu$ in Eq.~(\ref{rcirc2}), requiring only a modest $\lambda \gtrsim$ few to render $r_{\rm circ}/a \ll 0.5$. For instance, $\lambda \approx 3$ yields the constraint on the mass ratio $q \lesssim 36$. Irrespective of the actual balance of ram pressures in Eq.~(\ref{rambalance1}) below, the circularization constraint favors RBs ($q \lesssim 10$) over more extreme-mass-ratio $q \gtrsim 20$ BWs for the existence of a shock enshrouding the pulsar. If the shock $r_{\rm s}$ is constrained by other means, e.g., cooling breaks in hard X-rays, then an independent upper limit on $q$ is derivable. 

\subsection{Constraints on the Companion Mass Loss Rates}
\label{masscons}

The mass loss rate from the companion intrinsically couples to $r_{\rm s}$ and scenarios governing stability, as well as Eq.~(\ref{Breservoir1})-(\ref{Breservoir2}) for the magnetospheric scenario. Therefore, we briefly summarize constraints on the companion mass loss rates in BWs and RBs, which are generally much lower than that of Eddington-scale LMXBs. 

The existence of isolated recycled radio MSPs above the pulsar death-line suggests that time-averaged mass loss rates could be substantial, of order $-\langle\dot{m}_{\rm c} \rangle \sim 0.02 M_\odot$Gyr$^{-1}$ $\sim 10^{15}$ g s$^{-1}$, in many BWs and RBs if their evolutionary scenarios are similar. If $v_{\rm w} \sim v_{\rm K} \sim \sqrt{G M_{\rm p}/r_{\rm s}}$ near the shock and assuming the wind is gravitationally captured (cf. \S\ref{reductioadabsurdum}), then the condition $r_{\rm s} \gg r_{\rm circ}$ with $r_s \sim (\dot{m}_{\rm g} c^2/\dot{E}_{\rm SD})^2 R_g$ and $R_g = 2 G M_{\rm p}/c^2$ yields,
\begin{equation}
\dot{m}_{\rm g}  \ll \frac{\dot{E}_{\rm SD}}{v_{\rm orb} \, c} \sim 10^{17} \, \, {\rm g \, s^{-1}}
\label{mdotuplim}
\end{equation}
not an implausible bound for typical RB parameters. Here, we define mass rate participating in the shock $\dot{m}_{\rm g} \lesssim |\dot{m_c}|$ as a non-negligible fraction of the total companion mass loss rate $ |\dot{m_c}|$. Alternatively, one can constrain the total mass loss rate energetics of evaporation, $ \dot{m}_{\rm g} v_{\rm w}^2 \lesssim |\dot{m}_{\rm c}| v_{\rm esc}^2 \ll \Omega_{\rm msp} \dot{E}_{\rm SD}/(4 \pi)$ \citep{1988Natur.334..227V,1992MNRAS.254P..19S} where $\Omega_{\rm msp} \approx (0.5R_{\rm c}/a)^2$ is the solid angle fraction of pulsar wind intercepted by the companion. For typical RB parameters this yields the upper bound $\dot{m}_{\rm g} < |\dot{m}_{\rm c}| \ll 10^{18}$ g s$^{-1}$. 

 Additionally, from the form of the total binary angular momentum $J = M_{\rm p} m_{\rm c} \sqrt{a G/(  M_{\rm p} + m_{\rm c})}$ and Kepler III, one may show that in the no-accretion shock scenario when $\dot{M}_{\rm p} \approx 0$,
 \begin{equation}
 \frac{\dot{P}_b}{P_b} = 3\frac{\dot{J}}{J} + \frac{(-\dot{m}_{\rm c})}{m_{\rm c}}\left( 3 - \frac{1}{1+q} \right)
 \label{PbPbdot}
 \end{equation}
 for idealized point masses \citep{1924MNRAS..85....2J}. Measurement of orbital period derivatives in BWs and RBs by timing the MSP pulsations in the radio or $\gamma$-rays yield erratic and often negative values of order $ |\dot{P}_{b}| /P_{b} \lesssim 10^{-15}$ s$^{-1}$ rather than secular changes expected from conservative ($\dot{J}=0$) mass loss. The dominance of these nonsecular changes is interpreted in the \cite{1992ApJ...385..621A} framework, with the companion's gravitational quadrupole moment changing due to a magnetically active convection in the companion outer layers or activity cycles, with a significant portion ($\sim 10\%$) of the companion mass possibly asynchronous. There is some evidence for such changing gravitational quadruple moments in B1957+20 \citep{1994ApJ...436..312A}, BW J2051-0827 \citep{1998ApJ...499L.183S,2001A&A...379..579D,2011MNRAS.414.3134L,2016MNRAS.462.1029S}, RB J2339-0533 \citep{2015ApJ...807...18P}, and other MSPs \citep{2018ApJS..235...37A} suggesting that mass loss is lower than the simple Jeans formulation $|\dot{m}_{\rm c}| \lesssim m_c |\dot{P}_{b} |/P_{b} \sim 10^{17}$ g s$^{-1}$. Anisotropic pulsar emission, well-motivated theoretically and observationally in the framework of offset dipoles \citep[e.g.,][]{1996A&AS..120C..49A, 2011ApJ...743..181H,2015ApJ...807..130V,2016ApJ...832..107B}, may also cause quasi-cyclic wandering of $|\dot{P}_{b} |/P_{b}$ residuals \citep[][Eq. (76)]{1975ApJ...201..447H}. Therefore, these measurements constitute an upper limit for the mass loss rate.
   
Likewise, utilizing Eq.~(\ref{rochepac}) and again imposing $\dot{M}_{\rm p} = 0$, one may show that,
\begin{equation}
\frac{1}{2}\frac{\dot{R}_{\rm Lv}}{R_{\rm Lv}} = \frac{\dot{J}}{J} + \frac{(-\dot{m}_{\rm c})}{3 m_{\rm c}} \left( \frac{5}{2} - \frac{1}{1+q} \right).
\end{equation}
{If $\dot{J} < 0$ and since $-\dot{m}_{\rm c} \geq 0$ there exists a critical mass loss rate such that $\dot{R}_{\rm Lv} =0$, i.e. where the Roche potential radius switches between expansion and contraction. For a companion nearly filling its Roche lobe, a contracting Roche potential $\dot{R}_{\rm Lv}< 0 $ will drive mass loss towards the critical rate. Contrastingly, Roche radius expansion is only tenable via irradiation or ablation-driven mass loss beyond the critical rate when $\dot{J} < 0$. Secular gravitational wave angular momentum loss, $\dot{J}/J \approx  -32 G^3 M_{\rm p}^3(1+q)/(5 c^5 a^4 q^2)$ \citep{1975ctf..book.....L} specifies a minimum critical mass loss rate $|\dot{m}_{\rm c, crit}^{\rm GW}|$,}
\begin{eqnarray}
&&|\dot{m}_{\rm c, crit}^{\rm GW}| \approx \frac{192 G^3 M_{\rm p}^4 (1+q)^2 }{5 c^5 a^4 q^3 (3+5q)} \\
&& \sim 2 \times 10^{15} \left( \frac{M_{\rm p}}{1.7 \, M_\odot} \right)^4 \left( \frac{10^{11} \, \rm cm}{a} \right)^4 \left(\frac{7}{q} \right)^2   \, \, {\rm g \, s^{-1}} \nonumber
\end{eqnarray}
for $q \gg 1$, similar to the time-averaged $0.02 M_\odot$Gyr$^{-1}$ evaporative rate. 

Finally, a rudimentary lower limit may be estimated from radio eclipses of the radio MSP in BWs and RBs. After correcting for interstellar dispersion at uneclipsed orbital phases, excess delays near pulsar superior conjunction consistent with plasma dispersion generally imply the average dispersive free electron column density rises sharply from $ \langle n_e \rangle d \sim \Delta {\rm DM} \sim 10^{15}$ to $\gtrsim 10^{18}$ cm$^{-2}$, before total loss of radio emission in the eclipse \citep[e.g.][]{1991ApJ...380..557R, 2001MNRAS.321..576S,2009Sci...324.1411A,2013arXiv1311.5161A,2018MNRAS.476.1968P,2018Natur.557..522M,2018JPhCS.956a2004M}, for $d$ the line-of-sight distance through the plasma. In the absence of any clumping, e.g.,~at the shock, and $d \lesssim a$, this implies an isotropic mass loss rate $ |\dot{m}_{\rm c} | \gtrsim 4 \pi R_{\rm c}^2 (\Delta {\rm DM}) a^{-1} m_{\rm p} v_{\rm w} {\cal X}^{-1}$ for an ionization fraction ${\cal X}$. That is,
\begin{align}
|\dot{m}_{\rm c} | \, \gtrsim \, & 10^{13} \, {\cal X}^{-1} \sqrt{{\cal F}^3 \lambda} \left( \frac{\Delta {\rm DM}}{2\times 10^{18} \, \rm cm^{-2}} \right) \nonumber \\
& \times  \left( \frac{M_{\rm p}}{1.7 \, M_\odot} \right)^{2/3}  \left(\frac{P_b}{2 \times 10^4 \, \rm s} \right)^{1/3} \, \rm  g \, s^{-1}.
\end{align}
If the eclipse radius, which is a significant fraction of $a$, is utilized rather than $R_{\rm c}$, then the bound for $|\dot{m}_{\rm c} |$ is larger by a factor $(a/R_{\rm c})^2 \sim 10$ \citep{1994ApJ...422..304T}. Long-term variations of the deepness of eclipses may be used as a proxy for variations in ${\cal X}$ or the mass loss rate.

\subsection{Gravitational Influences and Wind Angular Momentum Loss}
\label{reductioadabsurdum}

As we discuss below, angular momentum loss of the companion wind is energetically essential for the ICDP shock state. The locale of such wind angular momentum loss is unknown. If it transpires far from the companion, the circularization radius constraints of \S\ref{circradsec} on the wind remain pertinent.

The stagnation point $r_{\rm s}$ balancing the ram pressure of the isotropic and supersonic two-wind interaction is given by
\begin{equation}
\frac{\dot{E}_{\rm SD}}{4 \pi c \,  r_{\rm s}^2} = \frac{|\dot{m}_{\rm c}| v_{\rm w}}{4 \pi (a- r_{\rm s})^2}.
\label{rambalance1}
\end{equation}
This implies the well-known stagnation point formula in terms of the ratio of wind ram pressures \citep{1990ApJ...358..561H},
\begin{equation}
\frac{r_{\rm s}}{a} = \frac{\sqrt{A_{\rm w}}}{1+\sqrt{ A_{\rm w}}} \quad, \quad A_{\rm w} \equiv \frac{\dot{E}_{\rm SD}/c}{|\dot{m}_{\rm c}| v_{\rm w}} \, .
\label{etawind}
\end{equation}
Anisotropic winds, as in \S\ref{anisoPwind}, modify these expressions but not the following general conclusions which are pertinent to the shock nose. Using Eq.~(\ref{etawind}) at $A_{\rm w} \leq 1$ or $r_{\rm s}/a \leq 0.5$, corresponding to the threshold of the shock orientation enshrouding the pulsar rather than the secondary, yields a lower limit on $v_{\rm w}$,
\begin{equation}
v_{\rm w} \gtrsim 10^{8.5} \left( \frac{ \dot{E}_{\rm SD} }{10^{35} \rm \, \,  erg \, s^{-1}} \right) \left( \frac{10^{16} \rm \,\, g \, s^{-1}}{|\dot{m}_{\rm c}|}  \right) \, {\rm cm \, s^{-1}} \, ,
\label{vws}
\end{equation}
exceeding the typical evaporative $v_{\rm esc}$ anticipated from an RB by at least an order of magnitude. Moreover, any lower $|\dot{m}_{\rm c}|$ than the high value used above yields untenably larger $v_{\rm w}$ values. There are also issues with energetics, with Eq.~(\ref{vws}) implying $\dot{E}_{\rm SD} \sim |\dot{m}_{\rm c}| v_{\rm w}^2$, clearly unjustifiable in an MSP self-excited wind scenario. Therefore the companion wind requires an additional reservoir of energy to tap in order for the shock to wrap around the pulsar. This contrasts the situation of high-mass pulsar X-ray binaries where the massive companion wind readily dominates the energetics. 

In the $\beta \gg 1$ scenario, a resolution to the apparent contradiction of Eq.~(\ref{vws}) is the influence of gravity of the MSP and angular momentum losses of the companion wind near or upstream of the shock. Two effects scale the ram pressure $\rho v_{\rm w}^2$ in quasi-spherical radial infall: a density enhancement nearer to the pulsar and Keplerian scaling of the fluid speed $v_{\rm K} \sim \sqrt{G M_{\rm p}/r_{\rm s}}$ for the gravitationally-influenced mass rate participating in the shock $\dot{m}_{\rm g}v_{\rm K} /(4 \pi r_{\rm s}^2)  \approx \rho v_{\rm K}^2$. Accordingly by pressure balance, the shock stand-off scales as $r_s/R_g \sim (\dot{m}_{\rm g} c^2/\dot{E}_{\rm SD})^2 \gg 1 $.

{Viscosity and heating of the companion wind somewhere within the pulsar Roche lobe (with $ r_{\rm circ} \ll r_{\rm s}$) is a critical requirement for angular momentum losses in the flow. As in accretion disks, turbulent viscosity is a possible dissipative mechanism. The viscous timescale then ought to be comparable to the free-fall dynamical timescale $\tau_{\rm ff} \sim P_b \sim 10^4$ s. This implies a kinematic viscosity $\nu_{\rm kin}$ of order $a^2/\tau_{\rm ff} \sim \nu_{\rm kin} \lesssim 10^{16}-10^{18}$ cm$^2$ s$^{-1}$, a rather high value that mandates hot ions with $c_{\rm s} \sim v_{\rm K}$. As we confirm in \S\ref{ADAF}, cooling of the flow is inefficient assuming such heating, which also is consistent with the $\beta \gg 1$ assumption. Therefore the requisite accretion flow solution must take the form of a quasi-radial heating- and advection-dominated accretion flow.} 

\newpage
\subsubsection{Instability}
\label{instab}

Quasi-spherical radial infall on a pulsar is unstable on dynamical timescales for $r_{\rm s}$ outside $r_{\rm LC}$, as demonstrated below and also touched upon by \citet{2001ApJ...560L..71B} in the context of accretion. One may regard the highly dissipative region near the termination shock and stagnation point as a fiducial volume subjected to the wind ram pressures (or momentum fluxes) originating far from this fiducial region. The MHD pressure ${\cal P}$ from the pulsar scales as ${\cal P} \propto r^{-n}$ with $n\approx6$ or $n\approx2$ inside or far outside the light cylinder, respectively, while the companion wind ram pressure $\rho v_{\rm K}^2 \propto r^{-k}$ with the Keplerian speed $v_{\rm K} = \sqrt{G M_{\rm p}/r}$ and $k \geq 5/2$ with the equality for zero angular momentum radial flows, and rising to $k \approx 2.6$ for Keplerian $\alpha$ disks \citep{1973A&A....24..337S}. That is, the momentum flux of the wind is
\begin{equation}
\rho v_{\rm K}^2 \sim \frac{\dot{m}_{\rm g} v_{\rm K}}{r^2} \propto r^{-5/2}
\label{infallrho}
\end{equation}
when $\dot{m}_{\rm g}$ is independent of $r$. Although straightforward generalization to regimes of Keplerian disks is attainable, for simplicity we assume quasi-spherical radial infall with $k\approx 5/2$ the remainder of the $\beta \gg 1$ scenario. Up to an irrelevant overall normalization, the pseudo-potential ${\varphi}$ associated with momentum fluxes on the dissipative fiducial volume, with some constant $A>0$, fulfills
\begin{equation}
- \nabla {\varphi} \sim -\frac{\partial {\varphi}}{\partial r} (r) \, \hat{\boldsymbol{r}} \sim \left( A r^{-n} - r^{-k} \right) \hat{\boldsymbol{r}}.
\label{pseudopotential}
\end{equation}
From the routine integration of Eq.~(\ref{pseudopotential}), $\varphi(r)$ may be obtained. Flux balance $-\partial {\varphi}/ \partial r (r_{\rm s}) = A r_{\rm s}^{-n} - r_{\rm s}^{-k} \equiv 0$, which defines the stagnation point $r_{\rm s}$ local equilibrium, is dynamically stable if and only if $\partial^2 {\varphi}/ \partial r^2 (r_{\rm s}) > 0$. This is satisfied when $n-k > 0$ implying stability only near the light cylinder where the near-zone MSP magnetic pressure may contribute if $k \gtrsim 2$ \citep[cf.][]{2001ApJ...560L..71B} unless the matter infall momentum flux is self-regulated (see \S\ref{gencrit}). 

Local anisotropy of the pulsar wind in the intermediate radiative zone for nearly-aligned rotators may raise stable radii $r_{\rm s}$ to several tens of $r_{\rm LC}$ \citep{2005ApJ...620..390E} a result derived for a vacuum Deutsch solution \citep{1955AnAp...18....1D} but which probably also holds for force-free and dissipative MHD winds. This is expected from the transition region between near-zone and far-zone expansion of the fields at the light cylinder. But such a situation so close to the pulsar still likely results in a disk rather than a shock due to nonzero angular momentum of the infalling matter if $r_{\rm circ} \sim r_{\rm LC}$. Consequently, the shock likely exists on orbital scales rather than near $r_{\rm LC}$, a proposition also supported by the lack of cooling breaks observed by NuSTAR \citep{2014ApJ...791...77T,2017ApJ...839...80W} owing to regions closer to the MSP (higher toroidial magnetic field) stipulating shorter synchrotron electron cooling timescales. Hence, mechanisms are required to stabilize the shock against perturbations on the dynamical timescales.
 
 \subsubsection{Summary}
To summarize, for the $\beta \gg1$ scenario:
\begin{enumerate}[label=(\alph*)]
\itemsep0em
\item{The circularization radius for the evaporative or qRLOF wind launched from the companion must initially be small $r_{\rm circ}/a \lesssim 0.5$, suggesting the mass loss be supersonic with the wind Rossby number inertia- rather than Coriolis-dominated ($>1$). This constraint is less demanding for mass ratios closer to unity and yields an upper limit for $q$.}
\item{The companion wind must convert to an ADAF somewhere within the pulsar Roche lobe prior to the shock. The more dense the wind plasma, the more naturally such viscosity/dissipative influences arise, especially near the shock. However, higher mass loss rates are more demanding to sustain for an evaporative or qRLOF scenario with $r_{\rm circ}/{a} < 0.5$, in contrast to a subsonic RLOF. Therefore some fine tuning is obligatory but poorly understood particularly since qRLOF may be anisotropic with clumping or density enhancements in the vicinity of the line joining the two stars. Such regimes ought to exist owing to the observed rotation-accretion power state transitions.}
\item{Self-regulatory mechanisms (\S\ref{mech}) must exist to stabilize the shock, so that it persists for a timescale of at least a few years. This also suggests all ICDP redback systems may be transitional binaries.}
\end{enumerate}

{

\subsection{Advection-Dominated Accretion Flow In ICDP Systems}
\label{ADAF}

In \S\ref{reductioadabsurdum} we deduced wind angular momentum loss is crucial under the hypothesis that the shock bows around the MSP in ICDP systems. Prior to assessing stability mechanisms for the shock in \S\ref{mech}, we examine an ADAF-like scenario and its observational consequences.

 In this ADAF-like scenario, the inflow is quasi-spherical, infall speed of order the free-fall speed, radiatively inefficient, optically-thin and sub-Eddington in mass rate \citep{1994ApJ...428L..13N,1995ApJ...452..710N}. The tenability of an ADAF-like scenario requires high plasma kinematic viscosity to furnish a heating-dominated flow \citep[for reviews, cf.][]{1998tbha.conf..148N,2002apa..book.....F}{; this is a critical assumption but not unconventional in the context of low-luminosity accretion flows.} 

Ions are preferentially heated in the ADAF regime \citep{1982Natur.295...17R}.  A two-temperature plasma results if (weak) Coulomb collisions are the only electron-ion equilibration mechanism operating far from the shock \citep{1995ApJ...452..710N} since electrons cool more efficiently than ions. As in $\alpha$ disks, turbulent viscosity is presumed to be the source of viscous dissipation and angular momentum transport. The microphysical mechanism to produce such turbulent viscosity is more speculative, but plausibly results from an Alfv\'{e}nic cascade or kinetic phenomena \citep[e.g.,][]{1998ApJ...500..978Q}. Longer-wavelength Alfv\'{e}nic MHD waves, corresponding to lengthscales $r_{\rm LC}$ of the MSP striped wind, may exist in the heating-dominated flow conditional on how prolific dissipative and reconnection processes are in the plasma near the shock, and how they influence the turbulence cascade.

The Eddington mass rate scale is,
\begin{equation}
\dot{m}_{\rm edd} \equiv \frac{4 \pi G M_{\rm p}}{0.1 \kappa_{\rm es} c} \sim 2 \times 10^{18} \, \, \rm g \, s^{-1}
\end{equation}
where $\sigma_T/m_{\rm p} = \kappa_{\rm es} = 0.4$ cm$^2$ g$^{-1}$ is the opacity and we have assumed the customary efficiency of $0.1$ \citep{2002apa..book.....F}. From constraints in \S\ref{masscons}, the mass rates in the rotation-powered state are sub-Eddington by one to three orders of magnitude. 

{We may consider if $\dot{m}_{\rm c} \sim 10^{15}-10^{16}$ g s$^{-1}$ required for pressure balance at the shock is inconsistent with the assumption of an ADAF-like solution possessing high viscosity. We note that in ADAFs, there exists a critical mass rate $\dot{m}_{\rm crit}$ where cooling terms begin to dominate heating terms and the self-similar ADAF approximation no longer holds. Mass capture rates higher than this critical rate result in a thin disk or luminous hot accretion flow \citep{2001MNRAS.324..119Y,2003ApJ...594L..99Y} rather than quasi-spherical infall. \cite{1997ApJ...477..585M} derives $\dot{m}_{\rm crit} \approx 0.3 \alpha^2 \dot{m}_{\rm edd}$ in their Eq.~(52), which suggests that  $\dot{m}_{\rm crit} \gtrsim 10^{16}$ g s$^{-1}$ when $\alpha \gtrsim 0.13$. For the large radii of interest here, i.e., $r_{\rm s} > 10^3 R_{\rm g}$, bremsstrahlung cooling is dominant and the plasma probably has a single temperature \citep{1995ApJ...452..710N}. In this simpler limiting case, i.e.,  Eq.~(4.1) of \cite{1995ApJ...452..710N}, the critical rate scales as $(r_{\rm s}/R_{\rm g})^{-1/2}$ and is approximately $\dot{m}_{\rm crit} \sim 50 \alpha^2 (r_{\rm s}/R_{\rm g})^{-1/2} \dot{m}_{\rm edd}$. This yields $\dot{m}_{\rm crit} \gtrsim 10^{-1.8} \dot{m}_{\rm edd} \sim 3 \times 10^{16}$ g s$^{-1}$ when $r_{\rm s}/R_{\rm g} \lesssim 10^5$ and $\alpha = 0.3$. Smaller radii further increase the critical rate limit. Therefore, we surmise that a disk does not necessarily result even at these higher mass rates provided that $\alpha \gtrsim 0.2$. 
}

In the following, we consider the ADAF scenario for RBs in a simplified approach, omitting a detailed balance analysis for the two-temperature plasma and neglecting factors of order unity. We assume a plasma number density set by radial infall (modulo a factor of $\sim 2$), 
\begin{equation}
n_{\rm tot} (r) \approx \frac{\dot{m}_{\rm g}}{2 \pi m_{\rm p} r^2 v_{\rm K} (r)} .
\label{ntot_simp}
\end{equation}

The ADAF is optically-thin to Thomson scattering when $r \ll 1/[n_{\rm e}(r) \sigma_{\rm T}]$, where $n_{\rm tot} \approx n_{\rm e}$. This is satisfied when 
\begin{equation}
r \gg \frac{\sigma_{\rm T}^2 \dot{m}_{\rm g}^2}{4 G \pi^2 m_{\rm p}^2 M_{\rm p}} \sim 10^3 \left(\frac{\dot{m}_{\rm g}}{10^{16} \, \, \rm g \, s^{-1}} \right)^2 \quad \rm cm
\end{equation}
which implies the ADAF is always optically-thin beyond $r_{\rm LC}$. Therefore, in the context of the ADAF hypothesis, the system is in the optically-thin sub-Eddington regime.

In the ADAF scenario, ions are heated and virialized by an underlying plasma kinematic viscosity $\nu_{\rm kin} \sim a^2/\tau_{\rm ff} \lesssim 10^{16}-10^{18}$ cm$^2$ s$^{-1}$. In the standard $\alpha$ description $\nu_{\rm kin} \sim \alpha c_{\rm s} a$ which implies a virialized ion thermal speed $c_{\rm s} \sim a/(\alpha \tau_{\rm ff})$. The requisite virialized temperature of ions $k_b T_{\rm vir} \sim m_{\rm p} v_{\rm K}^2 \sim G m_{\rm p} M_{\rm p}/r$ is,
\begin{equation}
T_{\rm vir} \sim  10^8 \left(\frac{2\times10^{10} \, \, \rm cm}{r}\right) \alpha^{-2} \quad \rm K.
\end{equation}
For the purposes of this rudimentary analysis, we assume electrons that are at a fraction $t<1$ of the ion virial temperature $ T_{\rm e} = t  T_{\rm vir}$.

 Several potential electron energy loss mechanisms operate. Synchrotron cooling, even under the presumption of equipartition (i.e. $\beta \sim 1$), does not contribute significantly until $T \gtrsim 10^9$ K or small shock radii $r_{\rm s}/a \ll 0.2$ \citep[][Fig. 6c]{1995ApJ...452..710N} and therefore is not considered. Compton cooling and electron-ion bremsstrahlung are two additional electron cooling channels. The thermal Compton volumetric energy loss rate in the Thomson regime is
\begin{equation}
\epsilon_{\rm comp} \sim n_{\rm e}  \sigma_{\rm T} c \,  u_{\rm rad} \Theta_{\rm e} 
\end{equation}
where $\Theta_{\rm e} = k_b T_{\rm e}/ (m_{\rm e} c^2) \ll 1$ \citep{1999APh....10...47B}. We assume the photon energy density fractionally reprocessing the pulsar spin-down, $u_{\rm rad} \sim \eta  \dot{E}_{\rm SD}/(4 \pi c r^2)$ with $\eta \sim 10^{-4} - 10^{-2}$.  Likewise, the thermal electron-ion bremsstrahlung volumetric cooling rate is $\epsilon_{\rm ff} \approx 1.7 \times 10^{-27} n_{\rm e}^2 T_e^{1/2}$ erg cm$^{-3}$ s$^{-1}$ \citep{1979rpa..book.....R}. Electron-ion bremsstrahlung is the dominant electron energy loss mechanism provided that
\begin{equation}
r \gtrsim 4 \times 10^8 \sqrt{t} \alpha^{-1} \left( \frac{\eta \dot{E}_{\rm SD}}{10^{33} \, \, \rm erg \, s^{-1}}  \right) \left( \frac{\dot{m}_{\rm g}}{10^{16} \, \, \rm g \, s^{-1}} \right)^{-1} \, \, \rm cm.
\end{equation}
Therefore Compton cooling is negligible for shock radii $r_{\rm s} \sim 10^{10}$ cm when $\dot{m}_{\rm g} \gtrsim 10^{15}$ g s$^{-1}$ or $\eta \sqrt{t} \alpha^{-1} \ll 10^{-2}$. 

The local thermal bremsstrahlung cooling timescale $n_{\rm e} k_b T_{\rm vir} /\epsilon_{\rm ff}$ is greater than the local dynamical timescale $r/v_{\rm K}$ when
\begin{equation}
r \lesssim 10^{15}  \left(\frac{10^{16} \, \, \rm g \, s^{-1}}{\dot{m}_{\rm g}} \right)^2  \, t \alpha^{-2} \quad \rm cm.
\label{bremscooling}
\end{equation}
Therefore the accretion flow is radiatively inefficient on orbital scales provided that $t \alpha^{-2} \gtrsim 10^{-4}$ or $\dot{m}_{\rm g} \lesssim 10^{18}$ g s$^{-1}$ when $t \alpha^{-2} \sim 1$.

\begin{figure}
\plotone{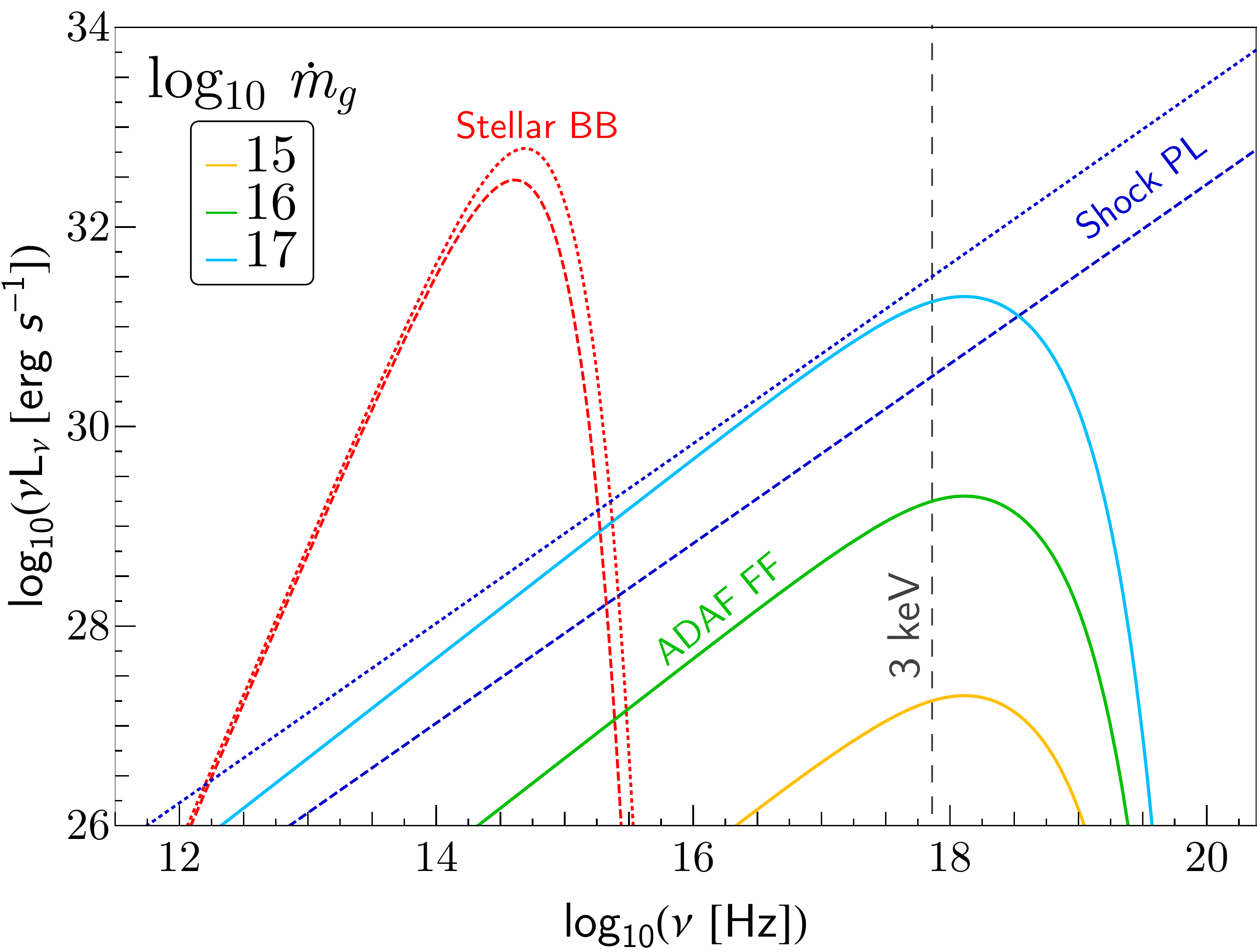}
\caption{Schematic SED from IR to 1 MeV of a representative ICDP system. Red and dark blue curves are illustrative of the thermal blackbody companion emission and shock power-law emission, respectively. The dotted and dashed curves bracket plausible extremes of orbital variation. The orbitally unmodulated thermal bremsstrahlung component is depicted with solid lines. See text for details. }
 \label{SED}
\end{figure}

The physical volume the ADAF occupies is uncertain and depends on the locales of heating {and spatial dependence of $\alpha$}. There are two limiting cases: a thin shell around the shock or the full spherical volume between the shock and companion $L_1$ point. In the latter, we may estimate the electron-ion bremsstrahlung spectrum $L_{\rm ff, \nu}$ for a virialized ADAF as the volume integral $L_{\rm ff,\nu} \sim 2 \pi \int_{r_{\rm s}}^{a} \epsilon_{\rm ff,\nu} r^2 dr$. Here $\epsilon_{\rm ff,\nu} \approx 8 \times 10^{-38} n_{\rm e}^2 T_e^{1/2} \exp{[- h \nu/ (k_b T_e)]}$ erg cm$^{-3}$ s$^{-1}$ Hz$^{-1}$ is the standard electron-ion bremsstrahlung volume emissivity \citep{1979rpa..book.....R} and we assume $t \alpha^{-2}$ is weakly spatially dependent. Then, the cumulative spectrum is
\begin{eqnarray}
L_{\rm ff, \nu} &\sim& 5 \times 10^{20} \,  \nu^{-1/2} \left( \frac{\dot{m}_{\rm g}}{10^{16} \, \, \rm g \, s^{-1}} \right)^2 \nonumber \\
&\times& \left( \erf[{\cal{S}}_{\nu}(a)] - \erf[{\cal{S}}_{\nu}(r_{\rm s})] \right)  \, \, \,  \rm erg \, \, s^{-1} \, Hz^{-1}
\label{bremspec}
\end{eqnarray}
where $\erf$ is the error function and
\begin{equation}
{\cal{S}_\nu}(r) = \left( t \alpha^{-2} \frac{G M_{\rm p} m_{\rm p}}{h \nu \, r} \right)^{-1/2}.
\end{equation}
At low energies, the spectrum is independent of $t \alpha^{-2}$ and flat in $\nu$. The spectrum breaks when ${\cal{S}_\nu}(a) \sim 1$, corresponding to a break energy,
\begin{equation}
\varepsilon_{\rm ff, b} \sim  \frac{G M_{\rm p} m_{\rm p}}{a} \, t \alpha^{-2} \approx 2.4 \,  \left(\frac{10^{11} \, \, \rm cm}{a} \right) t \alpha^{-2} \quad \rm keV.
\label{bremsbreak}
\end{equation}
A measurement of the break energy Eq.~(\ref{bremsbreak}) allows for an estimate of $t \alpha^{-2}$. 

The total electron-ion bremsstrahlung luminosity $L_{\rm ff}$ for a virialized ADAF is estimated by the volume integral $L_{\rm ff} \sim 2 \pi \int_{r_{\rm s}}^{a} \epsilon_{\rm ff} r^2 dr$,
\begin{eqnarray}
L_{\rm ff} \, &\sim& \, 6 \times 10^{29}  \sqrt{t} \alpha^{-1} \left( \frac{\dot{m}_{\rm g}}{10^{16} \, \, \rm g \, s^{-1}} \right)^2 \left( \frac{10^{11} \, \, \rm cm}{a} \right)^{1/2} \nonumber \\
 &\times& \left( \frac{1.7  \, M_\odot}{M_{\rm p}} \right)^{1/2} \left[ \sqrt{\frac{1}{r_{\rm s}/a} } - 1\right]\quad \rm erg \, \, s^{-1} .
\end{eqnarray}
The luminosity is below observational sensitivity for typical kiloparsec sources unless $\dot{m}_{\rm g} \gtrsim 10^{16}$ g s$^{-1}$ or $r_{\rm s}/a \ll 1$ when $\sqrt{t} \alpha^{-1} \sim 1$. A schematic SED for a typical ICDP system (see Table~\ref{RBenergetics}), spanning the infrared to MeV energies, is depicted in Figure~\ref{SED}. Red curves render idealized blackbody spectra for a companion of radius $R_{\rm c} = 3 \times 10^{10}$ cm and uniform temperature $T = 5000$ K (dashed) and $T = 6000$ K (dotted). Likewise, the non-thermal power law with photon index $\Gamma_{\rm Xs} = 1.1$ attributed to synchrotron emission from the intrabinary shock is depicted by the blue curves -- these are normalized in the $0.3-10$ keV band to $10^{31}$ erg s$^{-1}$ (dashed) and $10^{32}$ erg s$^{-1}$ (dotted). One may loosely interpret the dotted and dashed lines as bracketing pulsar inferior and superior conjunction phased-resolved luminosities, respectively. The orbitally unmodulated thermal bremsstrahlung component, via Eq.~(\ref{bremspec}) with $t \alpha^{-2}=1$, $a=10^{11}$ cm, $r_{\rm s}/a = 0.2$ and $M_{\rm p} = 1.7 M_\odot$, is depicted with solid lines for a range of $\dot{m}_{\rm g}$ values in g s$^{-1}$. 

For untenably large $\dot{m}_{\rm g} \gtrsim 10^{16.5}$ g s$^{-1}$, the bremsstrahlung emission may dominate the synchrotron power law at low energies and orbital phases near pulsar superior conjunction. {Yet, lower mass loss rates may also produce a detectable signal if there is significant clumping of matter in the infall (since the above rudimentary estimates assume isotropy), as suggested by the frequency dependence of radio eclipses.} This bremsstrahlung emission component, due to its turbulent nature, may exhibit stochastic time variability but such variations should be uncorrelated with orbital phase. Since such variability arises from the hydrodynamic nature of the flow, a red-noise character of variability is expected. Contrastingly, the synchrotron intrabinary shock emission ought to modulate with orbital phase, enabling discrimination between these components. Spectropolarimetry measurements offer additional discriminatory power. Even if $t \alpha^{-2}$ is much less than unity which impacts the cut-off energy Eq.~(\ref{bremsbreak}), the bremsstrahlung emission may be influential in the UV band. Indications of UV excess have been reported by \cite{2017arXiv170807041R} in J1227-4853 in its MSP rotation-powered state, but its origin is unclear. Future studies, particularly orbital phase-resolved UV to soft X-ray studies, are necessary to discern if any ADAF-mediated emission component exists. 

}

\subsection{Stability Mechanisms}
\label{mech}

\subsubsection{General Criteria}
\label{gencrit}

Radial infall in Eq.~(\ref{infallrho}) does not satisfy $n - k > 0$ for stability that follows from Eq.~(\ref{pseudopotential}) far from the pulsar light cylinder. However, if the captured mass rate $ \dot{m}_{\rm g}$ is coupled to $r_{\rm s}$, defined by where the momentum fluxes balance $-\partial {\varphi}/ \partial r (r_{\rm s}) = 0$, then stability may be attained in the MSP radiation zone where $n\approx2$. {Contrastingly, when $d \log  \dot{m}_{\rm g}/ d \log  r_{\rm s} \geq 5/2$, the assumptions of Eq~(\ref{infallrho}) and what follows in \S\ref{instab} are no longer valid, since gravitational influences of the MSP are negated.} Then, in the scenario pertaining to where the shock enshrouds the MSP, we demand the {\it{a priori}} constraint
\begin{equation}
\frac{1}{2} < \frac{d\log \dot{m}_{\rm g}}{d\log r_{\rm s}} < \frac{5}{2}
\label{indexrequirement}
\end{equation}
such that the gravitationally-influenced self-regulatory wind ram pressure scales as $\rho v_{\rm K}^2 \propto r_{\rm s}^{-k^\prime}$ with $0<k^\prime<~n~=~2$ in the MSP radiation far zone. 

We also stipulate the persistence timescale $\tau_{\rm p} = r_{\rm s}/\dot{r}_{\rm s} \sim 10^8$~s for metastability of the shock stand-off against much shorter-timescale dynamical perturbations. For such metastability, the response or relaxation time of any stabilizing process must be shorter than the dynamical perturbation timescale. In the stable regime when $n -k >0$, the frequency of such perturbations is proportional to $\sqrt{n-k}$. Therefore stability may be realized for arbitrarily long mechanism-dependent response times as $\sqrt{n-k} \rightarrow 0$. We may also safely assume the pulsar wind flux $\dot{E}_{\rm SD}/c$ is unchanged on timescales $\tau_{\rm p}$. 

We express $ \dot{m}_{\rm g} = \zeta |\dot{m}_{\rm c}| $ for $0 < \zeta < 1$, parameterizing the mass fraction captured gravitationally and participating in the shock. Then
\begin{equation}
\frac{d\log \dot{m}_{\rm g}}{d\log r_{\rm s}} = \frac{\partial \log |\dot{m}_{\rm c}|}{\partial \log r_{\rm s}} + \frac{\partial \log \zeta}{\partial \log r_{\rm s}}.
\label{sumindicies}
\end{equation}
The two terms in Eq.~(\ref{sumindicies}) correspond to two disparate routes for self-regulation between the shock radius and $\dot{m}_{\rm g}$: wind angular momentum capture ($\partial \zeta/\partial r_{\rm s} \neq 0$) that modulates $\zeta$ independent of the mass loss rate of the companion, and irradiation feedback ($\partial \dot{m_{\rm c}}/\partial r_{\rm s} \neq 0$) that influences the companion mass loss. These regulatory mechanisms may operate concurrently, nonetheless it may be plausible that one may dominate. For simplicity, we now consider the terms of Eq.~(\ref{sumindicies}) in isolation for the remainder of this Section. {Note that for plausible values of mass loss from the companion to attain pressure balance at the putative shock, we anticipate $\zeta$ to be not appreciably smaller than unity.} 

\subsubsection{Fractional Capture of the Wind, $\partial \zeta/\partial r_{\rm s} \neq 0, \partial |\dot{m}_{\rm c}|/\partial r_{\rm s} = 0$}
\label{fraccap}

For a wind which fulfills $r_{\rm circ} \ll r_{\rm s}$, the locale where the plasma momentum exchange and heating occurs between the companion and shock is uncertain. Clearly, the fluid viscosity and heating will be higher near the shock. We may speculate that the relative size of the shock as seen by the companion wind may regulate $\partial \zeta/\partial r_{\rm s} \neq 0$ such that shock stability is attained. Heating necessary to form an ADAF may be sharply peaked in a thin-shell locale nearby the shock. Under these assumptions, the ADAF-mediated emission component may be lower in total luminosity with uncorrelated shock X-ray synchrotron and companion optical variability. However, such fractional capture stability is difficult to quantify satisfactorily at this stage, and therefore we defer it to future work.

\subsubsection{Irradiation Feedback on the Companion, $\partial \dot{|m_{\rm c}|}/\partial r_{\rm s} \neq 0$, $\partial \zeta/\partial r_{\rm s} = 0$}
\label{irradfeedbacksec}

\begin{figure}
\plotone{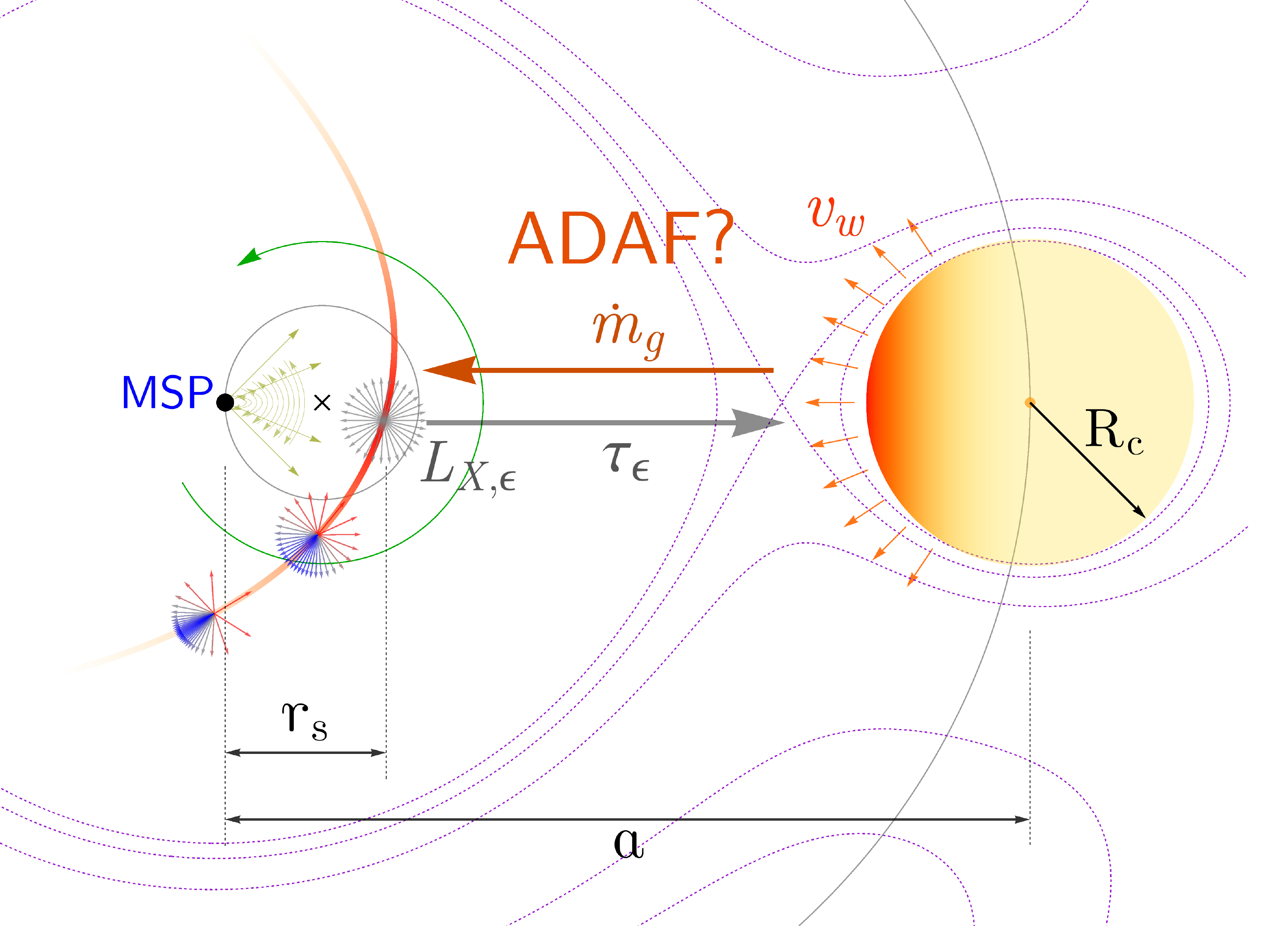}
\caption{Schematic face-on cross-sectional view of an RB system in the corotating frame in a state where the shock at $r_{\rm s}$ is bowed around the MSP with $q=4$ chosen for clarity.}
 \label{autoregcartoon}
\end{figure}

 Another natural mechanism for self-regulation is irradiation feedback on the companion from the shock and conversely. A dynamical equilibrium is established where shock irradiation modulates mass loss from the companion dependent on the shock location or radius $r_{\rm s}$, with larger (smaller) values of $r_{\rm s}$ corresponding to higher (lower) qRLOF mass loss from the companion. Systems close to the Roche limit in qRLOF may readily modulate their quasi-evaporative mass loss characteristics depending on a variety of influences, on a short timescale associated with perturbations in the pressure scale height of the outer atmosphere. Stability through irradiation feedback necessitates fast relaxation of the pressure scale height. {The short timescale of interest associated with such perturbations is $c_{\rm s}/g_{\rm c} \sim 10-100$ s $\ll P_{b}$, where $g_{\rm c} \lesssim 10^5$ cm s$^{-2}$ is the isolated-star surface gravitational acceleration and $c_{\rm s} \sim 10^6$ cm s$^{-1}$ the relevant photospheric sound speed.} Moreover, the induced wind information propagation timescale to the shock ought to be small compared to dynamical gravitational timescale that scales as $(n-k)^{-1/2}$.  
 
 We note that although $\gamma$-ray emission from the MSP magnetosphere may be the dominant {\textit{secular}} irradiating flux on the companion, the shock-mediated radiation (photons or particles) is assumed be the determinative nonsecular forcing that modulates mass loss on the companion. That is, secular $\gamma$-ray irradiation may result in a metastable companion state (at least on timescales $\sim \tau_{\rm p}$, cf. \S\ref{compinternal}) which is then further influenced by the shock irradiation. 
 
  As illustrated in Figure~\ref{autoregcartoon}, the dominant contribution of optically-thin shock emission that may influence the companion must originate from near the shock apex, owing to being the most radiatively-efficient region of the shock and Doppler-deboosting of emission in the wings away from the companion. From double-peaked X-ray light curves of such systems, this apex emission may be interpreted as the orbitally unmodulated X-ray background component of $L_{\rm X} \gtrsim 10^{31}-10^{33}$ erg s$^{-1}$. 
        
\begin{figure}
\plotone{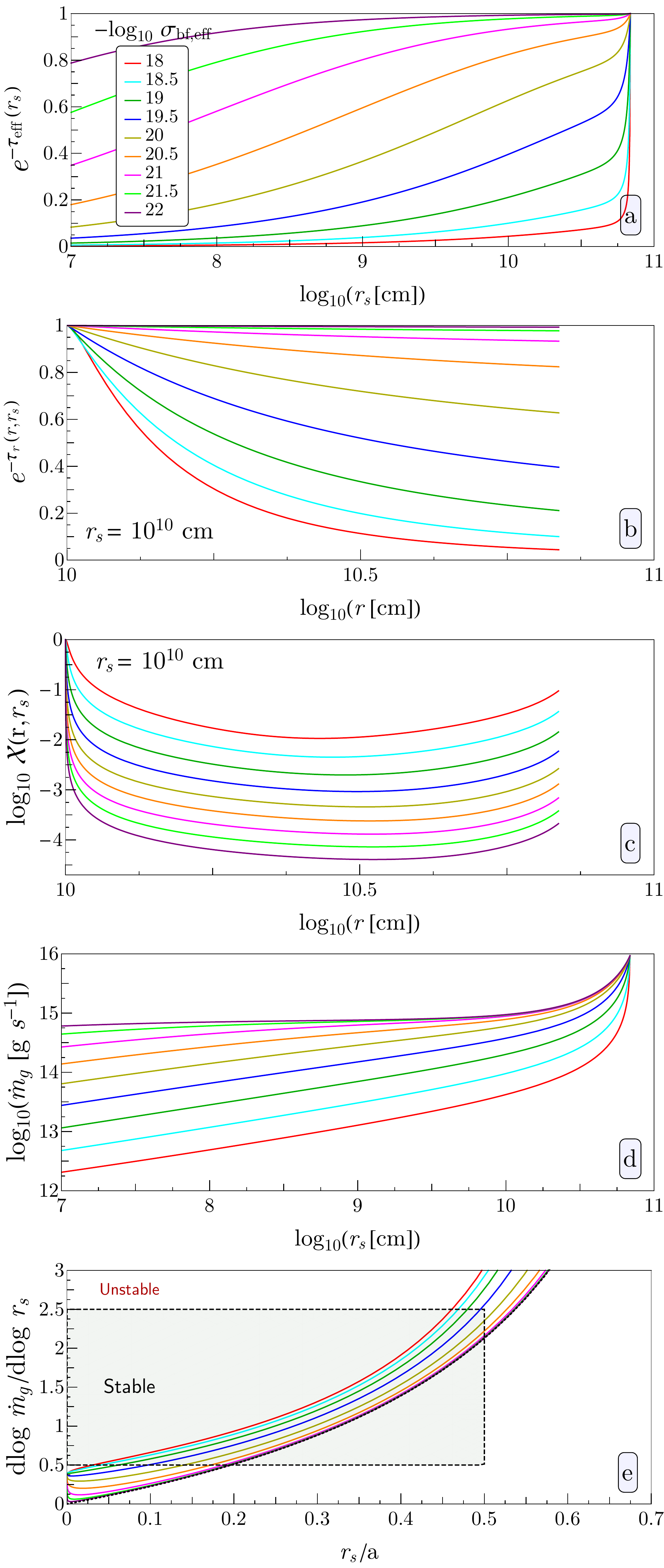}
\caption{Panel (a): Computation of the fixed-point optical depth $\tau_{\rm eff} = \tau_{\rm r}(a - L_1, r_{\rm s})$ as a function of shock stagnation point $r_{\rm s}$. Panel (b): The spatially-dependent optical depth $\tau_{\rm r}(a - L_1, r_{\rm s})$ at $r_{\rm s} = 10^{10}$ cm. Panel (c): Spatial dependence of ionization fraction with $r_{\rm s} = 10^{10}$ cm. Panel (d): Fixed-point mass loss rate from self-regulation. Panel (e): Stability region. Parameters: $\eta_{\rm \gamma} = 0$, $R_{\rm c} = R_{\rm vL}(q)$, $\eta_{\rm X} \dot{E}_{\rm SD} = 10^{32.5}$ erg s$^{-1}$, $q=7$, ${\cal{N}} = 0.5$, $M_{\rm p} = 1.7 M_{\odot}$, $a = 10^{11}$ cm, $T_{\rm c} = 6000$ K.}
 \label{pefigs_BF}
\end{figure}

\begin{figure}
\plotone{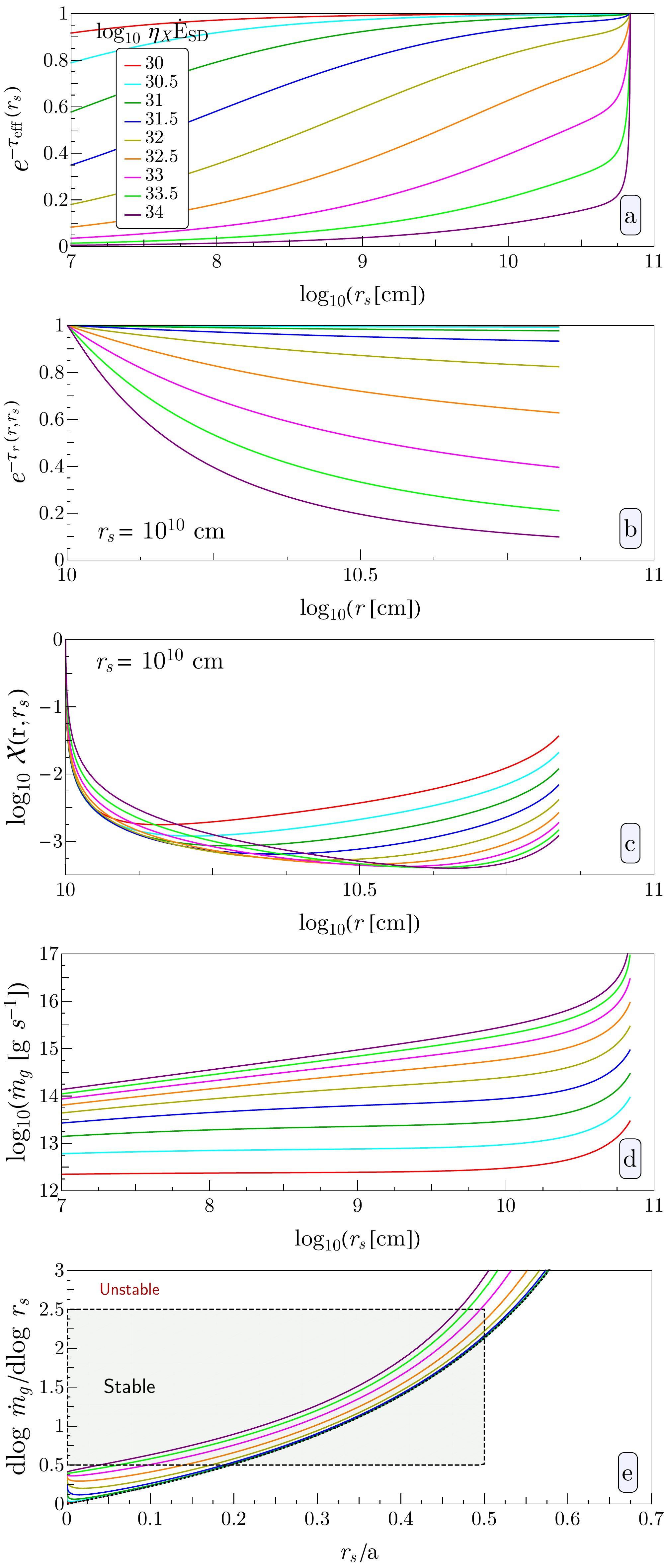}
\caption{Identical $\eta_{\rm \gamma} = 0$ construction as Figure~\ref{pefigs_BF} except with $\sigma_{\rm bf, eff} =10^{-20}$ cm$^2$ fixed with varying $\eta_{\rm X} \dot{E}_{\rm SD}$. }
 \label{pefigs_EDOT}
\end{figure}

\begin{figure}
\plotone{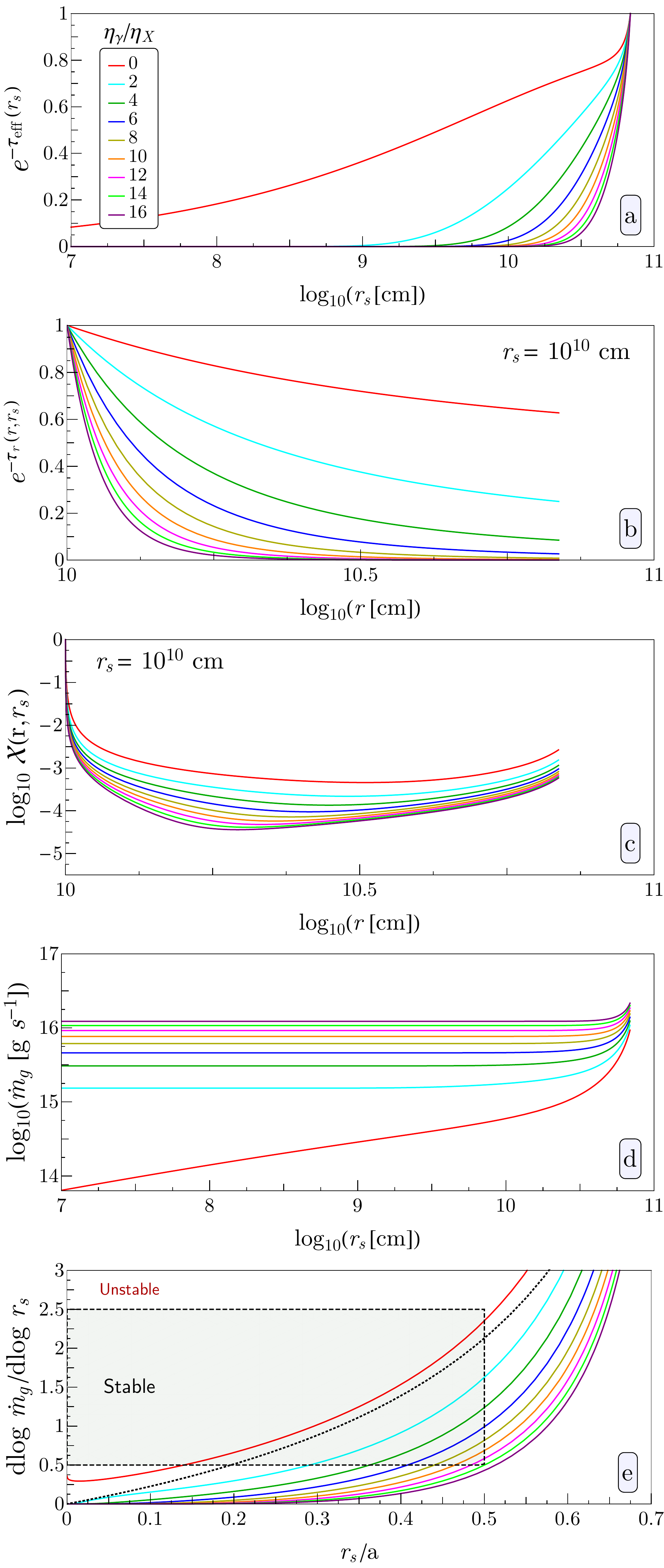}
\caption{Nonzero $\eta_{\rm \gamma}$ cases with $\dot{E}_{\rm SD} = 10^{35}$ erg s$^{-1}$, $\eta_{\rm X} = 10^{-2.5}$,  $\sigma_{\rm bf, eff} =10^{-20}$ cm$^2$, and otherwise the same parameters as Figure~\ref{pefigs_BF}.}
 \label{pefigs_ETAG}
\end{figure}

We now quantify the limit to the relative magnetospheric to shock radiative efficiencies permitted by stability. We assume that the companion mass loss rate scales linearly with impinging flux from the shock or MSP, parameterized by $\eta_{\rm X}$ and $\eta_{\rm \gamma}$, the energetic efficiencies from the shock emission and MSP magnetospheric emission, respectively. This convenient parameterization may also subsume particle heating contributions from the shock or MSP magnetosphere transported to the companion, but we emphasize the likely more dominant shock X-ray and MSP magnetospheric $\gamma$-ray emission with the subscript notation. The mass loss rate driven in irradiated stellar atmospheres is known to depend on the impinging power-law differential SED and is particularly sensitive to the soft X-rays relevant to atomic line heating/cooling \citep{1977ApJ...215..276B,1981ApJ...243..970L,1982ApJ...258..260L,2017MNRAS.467.4161D}. Therefore emission from the shock apex may be particularly pivotal on a metastable companion. Nonetheless, the gross linear form persists in the more sophisticated analyses. The accurate normalization of this linear $\dot{E}_{\rm SD}$ dependence is not essential in the ensuing stability analysis and may be scaled by a dimensionless constant ${\cal{N}}$ to satisfy the pressure balance condition $-\partial {\varphi}/ \partial r (r_{\rm s}) = 0$. {Such an expedient linear form has also been employed in the past, for instance, Eq.~(3.30) in \cite{1989ApJ...343..292R}. We stress that this scaling is highly uncertain and model-dependent. For instance, variation of the heating-cooling factor in \cite{1993ApJ...410..281T} may modulate the mass loss rate by two orders of magnitude for a fixed irradiation flux. {Moreover, the mass loss rate in these 1D models \citep[e.g.,][]{1993ApJ...410..281T}  may also vary by orders of magnitude when scaling the Roche filling factor that essentially controls surface gravity and consequently the local escape speed. Model assumptions also need modification for companions close to Roche-filling ${\cal{F}} \sim 0.8-1$ as in RBs, e.g., by Eq.~(34) of \cite{1982ApJ...258..260L} for deviations from inverse square law gravity near the companion surface. Without such modifications, the mass loss is likely significantly underestimated for near Roche-filling companions. Such nuances are beyond the scope of this paper.} We consequently encapsulate this complexity into dimensionless scalings ${\cal{N}},  \eta_{\rm \gamma}$ and $ \eta_{\rm X}$,}
\begin{eqnarray}
\dot{m}_{\rm g} &=& \zeta |\dot{m}_{\rm c}| \sim \frac{{\cal{N}}}{4 \pi v_{\rm esc}^2} \left[ \eta_{\rm \gamma} \Omega_{\rm c} (0)  +\eta_{\rm X} \Omega_{\rm c} (r_{\rm s}) \right]  \dot{E}_{\rm SD}
\label{mdotirrsimp}
\end{eqnarray}
where
\begin{equation}
\frac{\Omega_{\rm c} (r)}{4 \pi} \approx \frac{1}{2}\left(1 - \sqrt{1 - \left[\frac{R_{\rm c}}{(a- r)}\right]^2}\right)
\label{omegac}
\end{equation}
with $R_{\rm c} \approx R_{\rm vL}(q)$, $\Omega_{\rm c}$ the approximate solid angle fraction of the companion from the emission point, either the shock nose $r=r_{\rm s}$ or the MSP position $r=0$. {For typical RBs where $q \sim 7$, the solid-angle fraction may attain relatively large $\Omega_{\rm c}/(4 \pi) \approx 0.01 - 0.2$ values depending on the shock location.}
The stability condition Eq.~(\ref{indexrequirement}) with $\eta_{\rm \gamma} = 0$ and $ \partial \zeta/\partial r_{\rm s} = \partial \eta_{\rm X}/\partial r_{\rm s}~=~0$ then yields a minimum stable value of $r_{\rm s}$, 
 \begin{equation}
 \frac{r_{\rm s}}{a} \gtrsim 0.2 
 \label{rminsimp}
 \end{equation}
depicted as the crossing of the black dotted curve at $d\log \dot{m}_{\rm g}/d \log r_{\rm s} = 1/2$ in the bottom panels of Figures~\ref{pefigs_BF}--\ref{pefigs_ETAG}. This corresponds to the limit where $\gamma$-ray irradiation does not drive mass loss, but may render a metastable companion qRLOF state. Larger nonzero values of $\eta_{\rm \gamma}$ increase this lower limit and other effects such as photoelectric absorption may lower it (cf. \S\ref{photoelectric}).  {For the limiting case where $ \eta_{\rm \gamma} \Omega_{\rm c} (0)  \ll \eta_{\rm X} \Omega_{\rm c} (r_{\rm s})$, $q \gg 1$, and $r_{\rm s}/a = 0.2$,
\begin{eqnarray}
\dot{m}_{\rm g} \, &\sim& \, 8 \times 10^{15} \,  {\cal{N}} \left(\frac{q}{1+q}\right) \left( \frac{\eta_{\rm X} \dot{E}_{\rm SD}}{10^{33} \, \rm erg \, s^{-1} } \right) \nonumber \\
&& \times \left(\frac{M_{\rm p}}{1.7 \, M_\odot} \right)^{-1} \left( \frac{a}{10^{11} \, \rm cm} \right)  \quad \rm g \, \, s^{-1}.
\end{eqnarray}
This estimate is plausible and consistent with constraints in \S\ref{masscons} even if scaling constants are adjusted by one to two orders of magnitude. In qRLOF, the surface gravitational potential $\Phi \sim v_{\rm esc}^2$ may plunge dramatically when the companion is nearly Roche-filling. Such uncertainty here is encapsulated by scaling ${\cal{N}}$, therefore, it is not difficult for the irradiated wind to be rather intense.}

Imposing $r_{\rm s} \lesssim 0.5 \,a$ or $r_{\rm s} \ll a - L_1$, the limiting ratio of pulsar $\gamma$-ray to shock X-ray efficiencies permitting stability may be obtained from the logarithmic derivative Eq.~(\ref{mdotirrsimp}) and associating that to the lower bound of stability $1/2$ in Eq.~(\ref{indexrequirement}),
\begin{eqnarray}
\frac{\eta_{\rm \gamma}}{\eta_{\rm X}} &\lesssim& 12 +  5.34 \left(\frac{1}{q}\right)^{2/3} + 3.66\left(\frac{1}{q}\right)^{4/3} \quad r_{\rm s} \lesssim 0.5 \, a  \nonumber \\
\frac{\eta_{\rm \gamma}}{\eta_{\rm X}} &\ll& \frac{36 \,q}{\sqrt{5}} \qquad \qquad r_{\rm s} \ll a - L_1.
\label{etalimit}
\end{eqnarray}
The mass ratio $q$ dependence above arises from $L_1(q)$ and the construction of Eq.~(\ref{omegac}) with $R_{\rm c} \approx R_{\rm vL}(q)$ expanded in leading order of $1/q$. Intuitively, the upper bounds on $\eta_{\rm \gamma}/\eta_{\rm X}$ in Eq.~(\ref{etalimit}) merely convey that the MSP's magnetospheric $\gamma$-ray and particle emissions dilutes the stabilizing ability of the shock in this model.

Since the outer magnetosphere $\gamma$-ray beam from the MSP is wide, the assumption of quasi-isotropic emission couples $\eta_{\rm \gamma}/\eta_{\rm X}$ to the observable $F_{\gamma}/F_{\rm X}$. This estimate is robust up to beaming factors of order unity in both the numerator and denominator where $F_{\rm X}$ and $F_{\gamma}$ are the phase-averaged shock and pulsed $\gamma$-ray fluxes from the system. That is, when particle heating of the companion is not dominant,
\begin{equation}
\frac{\eta_{\rm \gamma}}{\eta_{\rm X}} \sim \frac{F_{\gamma}}{F_{\rm X}}.
\label{couple}
\end{equation}
The limit Eq.~(\ref{etalimit}) is an absolute demarcation in the phase space of stability. That is, when $r_{\rm s}$ attains the upper bounds in Eq.~(\ref{etalimit}) and when $d\log \dot{m}_{\rm g}/d \log r_{\rm s} = 1/2$, we obtain a robust upper limit to $\eta_{\rm \gamma}/\eta_{\rm X}$ if $\partial \eta_{\rm X}/\partial r_{\rm s} \approx 0$ since stable $r_{\rm s}$ are likely well below these upper bounds (cf. panel (e) of Figure~\ref{pefigs_ETAG}). Consequently, we anticipate $F_{\gamma}/F_{\rm X} $ to also be well below $F_{\gamma}/F_{\rm X} \sim \eta_{\rm \gamma}/\eta_{\rm X} \lesssim 14 \ll 160$ for $q \sim 10$ from Eq.~(\ref{etalimit}) even if the unknown beaming of order unity conspire against the association Eq.~(\ref{couple}). 

Observe that $\eta_{\rm X} \sim 10^{-3}-10^{-1}$ embodies the energetic efficiency of the total nonthermal shock emission, which may extend well into the hard X-rays where its energetics may dominate. A value $\eta_{\rm X} \sim 10^{-1}$ is similar to the Crab PWNe total synchrotron efficiency. Remarkably {\it all}~ICDP RBs exhibit $\gamma$-ray efficiencies well below unity $\eta_{\rm \gamma} \sim 10^{-2}-10^{-1}$ with $\eta_{\rm \gamma} \dot{E}_{\rm SD} \sim 10^{33}-10^{34}$ erg s$^{-1}$ \citep{2017ApJ...836...68T} even though $\eta_{\rm \gamma}$ approaches unity for many other MSPs. In particular, RBs J1023+0038 in its rotation-powered epoch \citep{2014ApJ...790...39S,2014ApJ...791...77T}, J1723-2837 \citep{2014ApJ...781....6B,2014ApJ...781L..21H,2017ApJ...839..130K} and J2129--0429 \citep{2016AAS...22821921N, 2018ApJ...861...89A, 2018arXiv180601312K} satisfy Eq.~(\ref{etalimit}). We are not aware of other published NuSTAR observations of RBs, but if Eq.~(\ref{etalimit}) is satisfied universally for RBs exhibiting ICDP X-ray light curves, then this aspect is compelling evidence for stability by irradiation feedback. Conversely, a BW or RB violating Eq.~(\ref{etalimit}) contemporaneously with ICDP state X-ray orbital modulation would prove challenging to the irradiation feedback $\beta \gg 1$ paradigm.

Table~\ref{RBenergetics} lists the energetics of RBs with extant {\it Fermi} constraints and computes the minimum energy $\varepsilon_{\rm min,cut}$ for extension of an unbroken X-ray intrabinary shock synchrotron power-law to satisfy $F_{\gamma}/F_{\rm X} \sim \eta_{\rm \gamma}/\eta_{\rm X} \lesssim 14$ (three sources with {\it{NuSTAR}} observations currently satisfy the bound). 

In the following, we consider several nuances of irradiation feedback and their observational signatures. In particular, irradiation feedback implies correlated optical and X-ray variability which may also be evident to a greater extent when $F_{\gamma}/F_{\rm X}  \ll 1$. If flaring states or epochs of the companion modulate the mass loss rate, then these may be correlated with X-ray variability on similar timescales. In the optical, such variability may manifest in subtle changes of line ratios or widths, transient emission or absorption lines or more dramatic variability in color temperature or optical orbital modulation. Cross-correlating X-ray flux with such optical signatures is therefore crucial to uncovering timescales associated with the modes of the irradiation feedback mechanism. In fact, as recently suggested by \cite{2017arXiv170605467S}, such correlated X-ray-optical variability may also be exhibited in SCDP BWs if ducted particle heating regulates the shock position or integrity. {There are indications of correlated X-ray-optical flux variability in at least one or two likely RBs \citep{2017ApJ...844..150H,2018arXiv180900215C}. This strongly motivates further multiwavelength scrutiny of rotation-powered RBs to discern if such variability is consistent with the irradiation feedback model.}

\paragraph{Intrinsic Shock Particle Acceleration and Beaming}
\label{intrinsicshock}

The shock energetic efficiency may depend on $r_{\rm s}$, i.e. $\partial \eta_{\rm X}/\partial r_{\rm s} \neq 0$ in Eq.~(\ref{mdotirrsimp}). If this is a dominant effect over simple solid angle elements in Eq.~(\ref{mdotirrsimp}), then the bound Eq.~(\ref{etalimit}) should be amended.  Relativistic shock acceleration in oblique shocks and the coupled companion mass loss is a poorly understood and highly nontrivial problem, therefore simple quantitative predictions are not feasible for Eq.~(\ref{indexrequirement}). However, some qualitative predictions may be discriminated.

Foremost, the toroidal magnetic field of the pulsar wind drops as $1/r_{\rm s}$ influencing the energy loss rate of electrons, the maximum electron Lorentz factor and influences the transport of any particle heating. It is unclear how this aspect affects the total efficiency of shock synchrotron emission, since although the energy loss rate rises with smaller $r_{\rm s}$, the maximum Lorentz factor and power-law index likely change. The efficiency $\eta_{\rm X}(r_{\rm s})$ may not strongly depend on $r_{\rm s}$ owing to the remarkably-narrow range of nonthermal X-ray power-law indices in rotation-powered RBs (Table~\ref{RBenergetics}). That is, even though $\eta_{\rm X}$ may be vastly disparate across the population of ICDP RBs, particle acceleration which predicates the nonthermal synchrotron emission is similar across putatively different toroidal magnetic fields or $r_{\rm s}$ values in the population of systems. We remark that in a simple radiation reaction limited acceleration scenario for the maximum Lorentz factor of leptons, the radiative power remains unchanged with magnetic field if Compton losses are neglected. 

The total energetics of acceleration may also become less efficient at smaller $r_{\rm s}$ notwithstanding the radial dependence of the toroidal the magnetic field, possibly a consequence of the higher particle losses in a smaller volume or unfavorable shock obliquity \citep[e.g.,][]{2012ApJ...745...63S}. If the shock narrows with smaller $r_{\rm s}$, a total phase-averaged X-ray dimming with decreased X-ray peak separation may be evident with unchanged or steepening X-ray photon power-law index. Contrastingly, if shock acceleration is more efficient at smaller $r_{\rm s}$ and biases total energetics towards harder X-rays above efficient atomic line excitations, then a flattening of the power-law index should be correlated with decreased X-ray peak separation. Such spectral changes of the shock emission would be correlated with variability in photospheric line ratios on the dayside of the companion

Additionally, the shock emissivity may be more beamed for smaller distances from the pulsar, a consequence of either the geometric narrowing of the shock or higher local bulk Lorentz factors along the shock tangent which produces the double-peak modulation. This scenario would imply higher pulsed fractions and narrower peaks in the double-peak X-ray orbital modulation, irrespective of intrinsic acceleration or dimming of phase-averaged flux.

\paragraph{Photoelectric Absorption in the Stellar Wind}
\label{photoelectric}

Photoelectric absorption of soft X-rays in a partially ionized wind may also influence stability the shock since it is the dominant absorption process for soft X-rays produced at the shock apex. Such absorption reduces the lowest stable $r_{\rm s}$ to well below that of Eq.~(\ref{rminsimp}), as will be apparent in due course. This necessitates a mean free path of photoionization of order the binary separation $a$ and relatively cool dayside companion temperatures $< 10^4$ K found in many RBs. {This may be in tension with the ADAF hypothesis of \S\ref{ADAF}, however, spatial regions close to the companion and $L_1$ point are necessarily cooler with $T\sim 5000$-$7000$ K at the companion photosphere. A working hypothesis here is that only the region near the shock is highly ionized, with cooler locales prevailing near the companion that moderate absorption and the mass loss rate. The ADAF may operate only at locales close to the shock. The physics of the mechanism we propose here is encapsulated in the optical depth $\tau_{\rm eff}$ and where it arises is not a critical aspect of the model if it moderates the mass loss rate.} Since absorption proceeds between the shock and companion, it does not influence inferior conjunction phase-centered Doppler boosted X-ray light curves from the shock, especially for geometries far from edge-on (see Figure~\ref{autoregcartoon}). We assume here the intrinsic shock X-ray emission does not scale with $r_{\rm s}$ for simplicity. 

Energy-dependent anisotropic radiative transport in an optically-thick medium is a highly nontrivial and nonlinear problem, therefore we adopt a simple 1D model based on scaling laws to capture the essential quantitative behavior of feedback. We assume the companion is in a metastable state due to secular $\gamma$-ray irradiation, with X-ray irradiation from the shock principally modulating the mass loss rate at qRLOF.

 Since the cross section for bound-free transitions possesses a strong energy dependence $\sigma_{{\rm bf,} \epsilon} \propto \epsilon^{-7/2}$ \citep{1979rpa..book.....R}, only the lowest energies contribute significantly in the optical depth integral at an effective optical depth and energy $\epsilon_{\rm eff} \lesssim 1$ keV, $\int_{\epsilon_1}^{\epsilon_2} L_{X, \epsilon} e^{-\tau_\epsilon} d \epsilon \approx L_{\rm X} e^{-\tau_{\rm eff}}$. For our present purposes, we take both $\eta_{\rm X} \dot{E}_{\rm SD}$ and $\sigma_{\rm bf,eff}$ as constrained parameters and consider absorption only along the line joining the shock apex and companion $L_1$ point.

Assuming $ \sigma_{\rm bf,eff} \gtrsim 10^{-20}$ cm$^2$ corresponding to the absorption cross section at $\epsilon_{\rm eff}  \sim 0.15$ keV, the photoionization timescale $\tau_{\rm ion} \sim 4 \epsilon_{\rm eff} \pi r_{\rm s}^2/(L_{\rm X} \sigma_{\rm bf,eff}) \sim 10^1$ s is much less than the dynamical timescale, satisfying a necessary condition for stability, i.e. fast relaxation. The recombination timescale is large, therefore the steady-state ionization fraction ${\cal X}$ is unity near $r_{\rm s}$, forming an ionization front. 
To quantify the influence of such absorption, we introduce an $e^{-\tau_{\rm eff}}$ factor for the shock term in Eq.~(\ref{mdotirrsimp}) for the radiative transport,
\begin{eqnarray}
\dot{m}_{\rm g} &=& \zeta |\dot{m}_{\rm c}| \nonumber \\
&\sim& \frac{{\cal{N}}}{4 \pi v_{\rm esc}^2} \left[ \eta_{\rm \gamma} \Omega_{\rm c} (0)  +\eta_{\rm X} \Omega_{\rm c} (r_{\rm s}) e^{-\tau_{\rm eff}} \right]  \dot{E}_{\rm SD}.
\label{mdotirr}
\end{eqnarray}
The photoionization and radiative recombination equilibrium ionization fraction ${\cal X}$ is spatially dependent for a given $r_{\rm s}$,
\begin{equation}
F_{\rm ion}  \, \sigma_{\rm bf,eff} \, n_{\rm tot} (1- {\cal X}) \approx  \alpha_{\rm H} \, n_{\rm tot}^2 {\cal X}^2 \quad ,
\label{ionequil}
\end{equation}
where $\alpha_{\rm H} \approx 4 \times 10^{-12}$ cm$^3$ s$^{-1}$ is an assumed spatially-independent radiative recombination coefficient for $T=10^4$ K \citep{1979rpa..book.....R}, $n_{\rm H}$ the total number density assumed purely hydrogenic for simplicity,
\begin{equation}
n_{\rm tot} (r,r_{\rm s}, \tau_{\rm eff}) \approx \frac{\dot{m}_{\rm g}( r_{\rm s}, \tau_{\rm eff})}{2 \pi m_{\rm p} r^2 v_{\rm K} (r)} 
\label{ntot}
\end{equation}
where $\dot{m}_{\rm g}( r_{\rm s}, \tau_{\rm eff})$ is given by Eq.~(\ref{mdotirr}), and $F_{\rm ion}$ the ionization flux along the line joining the shock nose and companion $L_1$ point attenuated by $e^{- \tau_{\rm r}}$, 
\begin{eqnarray}
&&F_{\rm ion}(r,r_{\rm s}; \tau_{\rm r}) = \frac{\eta_{\rm X} \dot{E}_{\rm SD} e^{- \tau_{\rm r}}}{4 \pi (r - r_{\rm s})^2} + \frac{R_{\rm c}^2 \sigma_{\rm B} T_{c}^{4}}{(a-R_{\rm c} -r)^2}  \label{Fioneq} \\
 && \qquad \qquad \qquad r_{\rm s} < r < a - L_1 \nonumber.
\end{eqnarray}
The optical depth $\tau_{\rm r}(r,r_{\rm s})$ encapsulates the attenuation from the shock apex to an intervening point $r$. We define $ \tau_{\rm eff} \equiv \tau_{\rm r}(a - L_1, r_{\rm s})$, the cumulative or fixed-point optical depth to the $L_1$ point of the companion which enters in Eqs.~(\ref{mdotirr})--(\ref{ntot}). The second term in Eq.~(\ref{Fioneq}) corresponds to photoionization flux from the companion with $R_{\rm c} < L_1$. The optical depth $\tau_{\rm r}(r,r_{\rm s})$ takes the form,
\begin{eqnarray}
\tau_{\rm r}(r,r_{\rm s}) &\approx& \sigma_{\rm bf,eff} \int_{r_{\rm s}}^{r } dr^\prime
 n_{\rm tot}(r^\prime,r_{\rm s}, \tau_{\rm eff} ) \nonumber \\
 &\times& \left[1- {\cal X}(r^\prime,r_{\rm s}, \tau_{\rm eff}, \tau_{\rm r} ) \right] \label{taueffint} \,,
\end{eqnarray}
which contains dependences on both $\tau_{\rm eff}$ and $\tau_{\rm r}(r,r_{\rm s})$ on the right-hand side through Eqs.~(\ref{mdotirr})--(\ref{Fioneq}). Taking the derivative of Eq.~(\ref{taueffint}) with respect to $r$ results in a nonlinear delayed differential equation for $\tau_{\rm r}(r,r_{\rm s})$ at each $r_{\rm s}$ with initial condition $\tau_{\rm r}(0,r_{\rm s})=0$ and constraint $\tau_{\rm eff} = \tau_{\rm r}(a - L_1, r_{\rm s})$. We solve the equation adopting a $\tau_{\rm eff}$ guess and iterating by root solving until $\tau_{\rm eff} = \tau_{\rm r}(a - L_1, r_{\rm s})$ to yield a self-consistent fixed-point solution. This in turn may be substituted back into Eq.~(\ref{mdotirr})--(\ref{ionequil})  to yield the spatial dependence of $\dot{m}_{\rm g}$ or the self-regulated equilibrium ionization front ${\cal X}(r,r_{\rm s})$. The examination of the stability criterion Eq.~(\ref{indexrequirement}) routinely follows. The result of this computation for typical RB parameters, for various values of $\sigma_{\rm bf, eff}$ and $\eta_{\rm X} \dot{E}_{\rm SD}$ are depicted in Figures~\ref{pefigs_BF}--\ref{pefigs_ETAG}. In all three figures, the minimum stable attainable $r_{\rm s}$ can be substantially below that of Eq.~(\ref{rminsimp}) which does not include any attenuation ($\tau_{\rm eff} = 0$).

In Figure~\ref{pefigs_BF}, $\eta_{\rm \gamma} = 0$ and $\eta_{\rm X} \dot{E}_{\rm SD} = 10^{32.5}$ erg s$^{-1}$ are fixed while $\sigma_{\rm bf, eff}$ is varied over four decades. Panel (a), depicting the fixed-point optical depth, clearly exhibits a strong dependence on the choice of $\sigma_{\rm bf, eff}$, with lower values of the cross section corresponding to lower attenuation. This may be compared with the $\tau_{\rm r}(r,r_{\rm s})$ calculation in panel (b), which increases monotonically with $r$ with endpoint $\tau_{\rm eff}$. Panel (c) depicts the ionization fraction, which manifests a sharp (on a linear scale) ionization front near $r_{\rm s} = r = 10^{10}$ cm, and a relatively weak dependence on $r$ thereafter except near the $L_1$ point where a rise in ionization fraction occurs due to the companion term in Eq.~(\ref{Fioneq}). For the mass rate $\dot{m}_{\rm g}$ in panel (d), higher values of $\sigma_{\rm bf, eff}$ attenuate shock emission so irradiation-induced mass loss is low. Conversely, low values of $\sigma_{\rm bf, eff}$ result in high $\dot{m}_{\rm g}$ with a relatively weak dependence on $r_{\rm s}$ corresponding to $d \log \dot{m}_{\rm g}/d \log r_{\rm s} \approx 0$ for $r_{\rm s} \ll 10^{10}$ cm; this is a weakly self-regulated regime of Eqs.~(\ref{mdotirrsimp})--(\ref{rminsimp}). Also in panel (d), the curves converge to the same value of $\dot{m}_{g}$ at $r_{
\rm s} = a - L_1$ set by Eq.~(\ref{mdotirr}). Panel~(e) depicts the stability regime, governed by Eq.~(\ref{indexrequirement}). Stability is realized with $ 0.02 \lesssim r_{\rm s}/a \lesssim 0.5$ for the entire suite of $\sigma_{\rm bf, eff}$ values, although larger and smaller values of $\sigma_{\rm bf, eff}$ are bounded at larger and smaller  $r_{\rm s}$, respectively. 

Figure~\ref{pefigs_EDOT}, which varies $\eta_{\rm X} \dot{E}_{\rm SD}$ while keeping $\sigma_{\rm bf, eff}$ fixed, showcases similar characteristics to Figure~\ref{pefigs_BF} for $\tau_{\rm eff}$ and the range of allowed $r_{\rm s}$ by stability. The former is principally due to the construction Eq.~(\ref{taueffint}) when $\tau_{\rm eff} \lesssim$ few. The dissimilarity principally arises since $\eta_{\rm X} \dot{E}_{\rm SD} $ directly scales $\dot{m}_{\rm g}$ in panel (d), and actually results in lower ionization fraction for larger $\eta_{\rm X} \dot{E}_{\rm SD}$ in panel (c) due to larger $n_{\rm tot}$. As in Figure~\ref{pefigs_BF}, the allowed stability range for lower $r_{\rm s}$ corresponds to the parameter choice that minimizes ${\cal X}$ along the intervening medium. This is a general feature of the model, even though there are clearly many free parameters. Therefore, in this scenario, a transient boost in ionization fraction, such as from a strong companion magnetic flare, may precipitate a state transition to an AMXP if $r_{\rm s}/a \lesssim 0.2$ prior to such a flare.

We consider the effect of nonzero $\eta_{\rm \gamma}/\eta_{\rm X}$ in Figure~\ref{pefigs_ETAG}. The additional mass loss driven by $\gamma$-ray irradiation saturates the mass loss rate and diminishes the stabilizing influence of photoelectric absorption. This is clearly evident with the flat mass loss rate in panel (d) and smaller phase space of stability in panel (e), with higher values of $\eta_{\rm \gamma}/\eta_{\rm X}$ shifting the stability curves to the right and higher minimum $r_{\rm s}$. This is in good agreement with the more simple zero-absorption geometric derivation Eq.~(\ref{etalimit}) owing to larger $\gamma$-ray irradiation induced mass loss sapping the self-regulatory influence of absorption.

If parameters such as the X-ray shock efficiency, mass loss rate, and $r_{\rm s} $ can be constrained observationally, an independent, albeit crude, constraint of the neutron star mass can be ascertained in this scenario from stability considerations alone. This arises owing to the gravitational term in Eq.~(\ref{ntot}).

Observational signatures of such photoelectric regulation include orbital phase-dependent IR/optical emission line features, and cross-correlations of these with nonthermal X-ray variability. Absorption lines from the companion may exhibit at inferior conjunction of the pulsar. Photoelectric absorption of the weak polar cap thermal emission of the MSP may be evident in a phase-dependent manner for systems near edge-on. If a system with ICDP X-ray light curves is found with very high dayside companion temperatures corresponding to high ionization fractions, such as for some BWs, then the stabilizing influence of this scenario may be low.

\newpage
\subsection{Summary of the $\beta \gg 1$ Scenario}

In the gas-dominance scenario, the circularization radius $r_{\rm circ}$ of the companion wind must initially be small so that a shock exists rather than a disk (since there is no evidence of disks in the rotation-powered ICDP state). This $r_{\rm circ}$ constraint generally favors less extreme binary mass ratios of RBs. If the shock enshrouds the MSP from pressure balance, then by energetic arguments prolific viscous heating and angular momentum loss of the companion wind is mandated at some point upstream of the shock. We hypothesize that this necessitates an ADAF-like solution.

The shock and ADAF-like configuration is unstable on dynamical timescales $P_b$ unless a self-regulatory mechanism operates in the system. Violation of the stability condition Eq.~(\ref{indexrequirement}) is tantamount to a system transition to a disk or ejector state. Stability of the shock in the $\beta \gg 1$ scenario may be realized via two channels, either fractional capture of the wind or irradiation feedback on the companion. Mechanisms in each mode have distinct testable observational signatures, although they may operate concurrently:
 
 \begin{enumerate}
\item{Fractional capture of the companion wind generally makes no predictions for correlated variability between the X-ray shock emission characteristics and companion activity. Moreover, there is no bound on the $\gamma$-ray to shock X-ray emission ratio $F_{\rm \gamma}/F_{\rm X}$.}
\item{Irradiation feedback anticipates such correlated variability, and operates with different channels that couple the companion mass loss rate to the shock stagnation point location $r_{\rm s}$. The $\gamma$-ray efficiency of the MSP cannot be too large as to dilute the stabilizing ability of the irradiation feedback mechanism, with $\eta_{\rm \gamma}/\eta_{\rm X}$ satisfying the bound of Eq.~(\ref{etalimit}). This motivates soft and hard X-ray observations of ICDP-type BWs and RBs to ascertain whether this limit is universal.}
\begin{enumerate}
\item{The intrinsic shock acceleration and emissivity may strongly depend on distance from the MSP. This scenario may be discriminated by correlated X-ray spectral and companion variability, e.g., line ratios on the day side of the companion cross-correlated with nonthermal X-ray spectral changes.}
\item{The relative beaming of irradiation and associated bulk Lorentz factor in the shock may depend on $r_{\rm s}$. This may be exhibited by interdependence between properties of the double-peaked X-ray light curve morphology and companion variability.}
\item{The presence of photoelectric absorption of shock emission within the companion wind flow may be tested by orbital phase-dependent emission and absorption line features. Additionally, hard X-ray cooling breaks in ICDP systems which indicate a shock at $r_{\rm s} \ll 0.2 \, a$ may suggest the need for photoelectric absorption for stability.}
\end{enumerate}
\end{enumerate}

\section{Internal Companion Dynamics, the Irradiation Blanket Effect and Long-Term Instability}
\label{compinternal}

Instability, and therefore transitions to AMXPs may also be driven on much longer timescales by the internal dynamics of the companion which may precipitate changes in the mass loss rate, and therefore stability mechanisms in the $\beta \gg 1$ scenario or obviate the $\beta \ll 1$ assumption for the magnetospheric scenario (see Eq.~(\ref{Breservoir1})--(\ref{Breservoir2})).

Low-mass nondegenerate stars such as those in RBs are highly convection- rather than radiation-dominated in their internal energy transport. From entropy considerations for convective stellar structure, irradiation on such stars may inhibit internal energy transport on the dayside \citep{1985A&A...147..281V,1990A&A...228..231N}, the so-called blanket effect. Such an irradiation effect causes bloating of a convection-dominated companion.  In convective stars, mixing is the most efficient energy transport mechanism with subsonic mixing speeds $v_{\rm mix} \sim (L_{\rm c}\Delta R_{\rm c}/m_{\rm c})^{1/3} \sim 10^2-10^4$ cm s$^{-1}$ and associated timescales of order $\tau_{\rm mix} \sim \Delta R_{\rm c}/v_{\rm mix} \sim 10^5-10^8$ s for layer thickness $\Delta R_{\rm c}$ and intrinsic stellar luminosity $L_{\rm c}$. 

This $\tau_{\rm mix}$ is the associated timescale of the convective dynamo, and in the $\beta \ll 1$ scenario links to variability and stability of the magnetosphere between force-free equilibria, provided that the mass loss rate is low. In the $\beta \gg 1$ scenario, if irradiation feedback is a stabilizing mechanism, then deep mixing is uninvolved owing to being much longer than the dynamical timescale $\tau_{\rm ff}$. Equilibration to impulsive changes in the irradiation flux, e.g., state transitions between rotation-powered or accretion states, then occurs on the order of $\tau_{\rm mix}$ assuming such states persist for at least that long. This is particularly interesting in the case of an AMXP transitioning to a rotation-powered state where the lingering irradiation-induced mass loss may endure as qRLOF. 

For convective stars, anisotropic irradiation and the blanket effect alters the global scaling among metastable equilibrium luminosity $L_{\rm e,c}$, radius $R_{\rm c}$ and photospheric temperature $T_{\rm e,c}$. From the model of \cite{2000A&A...360..969R} Eqs.~(16)--(19), we establish
\begin{align}
\frac{\dot{L}_{\rm e,c}}{L_{\rm e,c}} \approx& -7.5 \frac{\dot{R}_{\rm c}}{R_{\rm c}} \\
\frac{\dot{T}_{\rm e,c}}{T_{\rm e,c}} \approx& \, \, 0.03 \frac{\dot{R}_{\rm c}}{R_{\rm c}} \, \approx \, - 0.004 \frac{\dot{L}_{\rm e,c}}{L_{\rm e,c}}.
\end{align}
\\
Since $\beta \gg 1$ in the interior, the influence of a dynamo is inconsequential for these relations. Therefore, the effective temperature is very weakly dependent while luminosity is strongly (and inversely) dependent on changes in the equilibrium radius, in contrast to simple Steffan-Boltzman scaling $L_{\rm e,c} \approx 4 \pi R_{\rm c}^2 \sigma_{\rm B} T_{\rm e,c}^4$. {The model of \cite{2000A&A...360..969R} attempts to account for the anisotropic irradiation and energy loss from the unirradiated stellar night side. The predictions are qualitatively different than simpler isotropic-irradiation models where no such energy loss mechanism is allowed. For instance, the companion bloating in anisotropic irradiation models is much weaker than in isotropic models \citep[cf. discussion in][and references therein]{2000A&A...360..969R}. }

 If the long-term optical dimming in J2129--0429 of $-\dot{L}_{\rm e,c}/L_{\rm e,c}\approx 8 \times 10^{-4}$ yr$^{-1} \approx (4\times 10^{10})^{-1}$ s$^{-1}$ \citep{2016ApJ...816...74B} is attributed to such outer convective envelope physics, then $\dot{T}_{\rm e,c}/T_{\rm e,c} \approx 10^{-13}$ s$^{-1}$. This small temperature change is consistent with no temperature change observed in J2129--0429 spectroscopically by \cite{2016ApJ...816...74B}. Moreover, this implies an {\textit{increasing}} stellar radius $\dot{R}_{\rm c}/R_{\rm c} \approx 3 \times 10^{-12} $ s$^{-1}$ or $\Delta R_{\rm c}/R_{\rm c} \sim 10^{-3}$ over a decade. Intriguingly, \cite{2018ApJ...861...89A} reported that J2129--0429 recently transitioned to brightening at a rate of about  $\dot{L}_{\rm e,c}/L_{\rm e,c}\approx 3 \times 10^{-3}$ yr$^{-1} \approx 10^{-10}$ s$^{-1}$ corresponding to a contraction rate $\dot{R}_{\rm c}/R_{\rm c} \approx - 10^{-11} $ s$^{-1}$, an order of magnitude larger than in the dimming phase. The limited cadence of the observations restrict the utility of any assessments of the brightening transition or second derivatives at this stage. If the Roche Lobe fraction ${\cal{F}} \approx 0.95$ for J2129--0429, then the transition time for a radius change of $5\%$  is of order $10^9-10^{10}$~s (i.e. to RLOF in the expansionary phase). This is much shorter than the thermal relaxation (Kelvin-Helmholtz) timescale due to irradiation by $\dot{E}_{\rm SD}$ with $R_{\rm c} \approx R_{\rm vL}$, $\tau_{\rm therm} \sim 20 G m_{\rm c}^2 (1+q)/(a\dot{E}_{\rm SD}) \sim 10^{14}-10^{15}$~s for typical RB parameters. Therefore, a long-term secular dimming, if sustained, may augur transitions to an AMXP in some RBs as mass loss or $r_{\rm circ}$ secularly increases. Observe that the timescale $10^{10}$~s is an upper limit since Eq.~(\ref{Breservoir1})--(\ref{Breservoir2}) or the stability criterion Eq.~(\ref{indexrequirement}) may be violated well before the onset of RLOF. Likewise, internal variability associated with convective mixing or irradiation feedback from the shock also operate on shorter timescales.
 
\section{Discussion and Observational Discriminants}
\label{summary}

For intrabinary shocks in low-mass MSP binaries where the putative shock configuration bows around the pulsar, the mechanism yielding pressure balance for such a configuration in the rotation-powered state is a crucial unresolved issue. We have examined two scenarios, intimately connected with the nature of the companion and its mass loss. We suggest that in either scenario, all ICDP RBs may be transitional systems. To overpower the pulsar wind any scenario requires somewhat extreme assumptions. We now compare the two scenarios and their distinguishing observables.

In the $\beta \ll 1$ scenario, a robust companion magnetosphere exists with poloidal surface fields of several kilogauss. This magnetospheric scenario's immediate stability is coupled to the global magnetospheric stability, i.e. the MHD force-free equilibria, which ought to persist at least as long as $\tau_{\rm p} \gtrsim 10^8$ s. The companion mass loss rate must be relatively low $|\dot{m}_{\rm c}| \lesssim 10^{15}$ g s$^{-1}$ for magnetic dominance to be sustained for $\tau_{\rm p}$. The relatively low mass loss rate and intrinsic stability of this scenario offers several virtues over the $\beta \gg 1$ scenario. Yet, there are a number of unresolved issues. Although RB companions are not isolated main sequence stars, it appears difficult to generate more than a few kilogauss fields in even rapidly-rotating M dwarfs \citep{2009ApJ...692..538R}. Such large fields imply that much of photosphere attains $\beta_{\rm c} \sim 1$, suggesting large starspots. Yet there is no definitive evidence of large star spots in most RB companions \citep[except perhaps J1723--2837,][]{2016ApJ...833L..12V} although there is some evidence of flares \citep{2016ApJ...823..105D, 2018arXiv180900215C} albeit not as extreme as in some BWs in the optical. Moreover, the orientation of the companion dipole moment is unknown, but certain orientations ought to strongly influence the magnetic obliquity of the shock, possibly suppressing conventional diffusive shock acceleration. Finally, it is unclear how gas dominance is acquired in the AMXP disk state for a sustained time of months or more, and how the transition from the rotation-powered $\beta \ll 1$ state to accretion transpires. Two possibilities are that the mass loss rate via RLOF overwhelms the magnetosphere, or the companion dynamo is substantially weakened by an unknown mechanism. The former scenario, however, then requires mass loss rates much higher than that inferred in disk states of J1227--4859 \citep{2015MNRAS.449L..26P} and J1023--0038 \citep{2015ApJ...807...33P}. One could propose that there is strong spatial stratification of $\beta$ such that $\beta \ll 1$ near the companion but which transitions to $\beta \gg 1$ in the pulsar Roche lobe. However, such a scenario is difficult to justify since for the magnetospheric field required for shock pressure balance, $\beta \ll 1$ everywhere for reasonable plasma densities and temperatures. Therefore we surmise that companion internal dynamics and the intrinsic dynamo mechanism play a crucial role in state transitions between pulsar and AMXP states in the magnetospheric scenario. The spin axis funnels in \S\ref{anisoPwind}, and whether they are connected with the main shock surface, may also be influential in state transitions and sporadic accretion and requires future study with MHD simulations. The topological changes associated with a time varying companion dynamo ought to manifest in strong reconnection flares with fluence bounded by $B_{\rm c}^2 R_{\rm c}^3 \lesssim 10^{37}$ erg, e.g., when the spin axis funnels join or separate from the principal bow shock surface, or reconnection events occur behind the pulsar.

Contrastingly, it is conceptually simpler in the $\beta \gg 1$ scenario to recognize why transitions to AMXP states may transpire since this scenario is inherently unstable without stability mechanisms. Substantial mass loss rates may be realized in the qRLOF regime, aided by strong irradiation from the pulsar and shock. RB companions observed to be close to Roche-lobe filling and exhibiting substantial radio eclipses of the MSP indicating the presence of plasma in the system is supportive of this picture. Yet, the mass loss rates required are substantial $|\dot{m}_{\rm c}| \sim 10^{15} - 10^{16}$ g s$^{-1}$, perhaps exceeding those rates in the low-luminosity disk states where disk truncation is inferred near the conventional Alfv\'{e}n radius $r_{\cal A} \sim 10^7 \left[\dot{m}_{\rm g}/(10^{15} \, \rm g \, s^{-1})\right]^{-2/7}$~cm proximate to $r_{\rm LC}$. The high mass loss rates pose a conceptual obstacle, yet are not ruled out by any observational constraints in \S\ref{masscons}. A related issue is that for the shock enshroud the MSP, then prolific viscous heating and angular momentum loss of the companion wind are mandated at some point upstream of the shock. We hypothesize that this necessitates an ADAF-like solution but the feasibility of the high requisite viscosity is questionable. Fortunately, as we established in \S\ref{ADAF}, bremsstrahlung emission from the ADAF flow does not result in a new SED component which would be in tension with any current observations. Finally, the question of stability is a serious issue for $\beta \gg 1$ scenario and demands stability mechanisms that may be diagnosed with monitoring observations.

How might one distinguish between the magnetosphere and gas-dominance scenarios? A direct measurement of a kilogauss magnetic field of the companion, for instance by Zeeman line splitting \citep[i.e.,][]{2012LRSP....9....1R}, would be highly suggestive of the $\beta \ll 1$ scenario but not conclusive since such fields may be localized to spots and do not constrain the mass loss. Moreover, for the magnetospheric scenario no ICDP RBs ought to be found with hot companion atmospheres where the dominant source of continuum opacity is bound/free-free emission [see Eq.~(\ref{plasmabetaK})]. Likewise, the $\beta \gg 1$ scenario may be immediately falsified if systems are found exhibiting ICDP X-ray orbital modulation with companions that cannot support high mass loss rates, for instance, by a companion that is significantly Roche-lobe underfilling or unusually hot and compact. However, for transitional systems with Roche-filling companions like J1023+0038 and J1227--4859, discrimination requires further investigation. A direct diagnostic is by careful study of radio eclipses of the pulsar, particularly for MSPs which have intrinsically strong polarization at inferior conjunction. Such radio spectropolarimetric study of eclipses proximate to ingress and egress directly probes the plasma in the system and proximate to the shock. Imaging radio studies of RBs suggest absorption rather than scattering for the nature of the eclipses \citep{2015ApJ...800L..12R,2016MNRAS.459.2681B} with cyclotron absorption in a strong magnetic field being often referenced, seemingly favoring the magnetospheric scenario. However, relativistic electrons mandated by the recent discoveries of the non-thermal orbitally-modulated X-ray emission also enable synchrotron absorption \citep{1991ApJ...370L..27E,1994ApJ...422..304T} as a potentially viable mechanism. Indeed, as detailed by \cite{1994ApJ...422..304T}'s Appendix C, synchrotron absorption can be particularly sensitive to the nature of the electron distribution function, particularly its nonthermal tail. Note that in the linear plasma limit, for the magnetospheric scenario there ought to be a rotation measure near the peripheries of an eclipse (where the medium is optically thin) of order,
\begin{align}
| \Delta {\rm RM} | \sim 2.6 \times 10^4 \left( \frac{| \langle B_\parallel \rangle |}{100 \, \rm G} \right) \left( \frac{\Delta {\rm DM}}{10^{15} \, \rm cm^{-2}} \right) \, \, \rm rad \, m^{-2}
\end{align}
where $\Delta {\rm RM}$ is an orbital-phase dependent rotation measure after interstellar corrections. Compelling evidence for the magnetospheric scenario would then be orbital phase dependence of such large $\Delta {\rm RM}$, with perhaps even changes in sign if the companion magnetic moment is skewed as in Figures~\ref{Aniso15angled}--\ref{Aniso90angled} or multipolar. This may be challenging, as the large Faraday rotation combined with dispersive delays may effectively lead to depolarization over longer integration times. Moreover, nonlinear plasma processes and lensing may be operating. Indeed, the $| \Delta {\rm RM} |$ is similar to that observed in some FRBs. As noted by \cite{1994ApJ...422..304T}, an even more effective probe is the polarization of individual pulses near eclipses, for instance, for diagnosing the mode propagation and pulse splitting \citep{2018arXiv180809471S}. Recent radio studies similar to that of eclipsing BWs such as \cite{2018MNRAS.476.1968P} and \cite{2018Natur.557..522M} applied to ICDP RBs would be particularly useful, not only as a diagnostic of the eclipse medium and magnetic field but also the eclipse mechanism.

Indirect evidence may also distinguish between the magnetosphere and gas-dominance scenarios. If the mass loss rate is high, the ADAF in the $\beta \gg 1$ scenario offers the prospect of stochastic (red-noise type) orbitally unmodulated thermal bremsstrahlung emission components in the UV and soft X-rays, which may dominate the shock synchrotron emission at orbital phases near pulsar superior conjunction (see Figure~\ref{SED}). Stability mechanisms such as irradiation feedback in the $\beta \gg 1$ scenario may also exhibit signatures, such as correlated X-ray-optical variability. Likewise, irradiation feedback mechanism's stability criterion implies $F_\gamma/F_{\rm X} \lesssim 14$ which may be tested on the current and future population of ICDP systems by probing for spectral cut-offs in the hard X-ray band. Finally, we note that an intriguing probe on the population of ICDP systems is the asymmetry of the double peaks in X-ray orbital modulation. The magnetospheric scenario has no strongly preferential asymmetry since the companion dipole moment may be oriented in any direction relative to the orbital plane, while in the $\beta \gg 1$ scenario Coriolis influences impart asymmetry of the leading and trailing peaks in the shock shape and thereby model light curves \citep{2016arXiv160603518R,2017ApJ...839...80W}. Therefore, future population studies of asymmetries in X-ray light curves may be insightful.  

Contemporaneous X-ray, IR/optical, and radio observations and long-term monitoring in the MSP state of RBs are therefore crucial to disentangle different modes and mechanisms for shock pressure balance, stability and elucidate the nature of transitional millisecond pulsar binaries. 

\acknowledgements
We thank the anonymous referee for insightful questions, comments and suggestions. Z.W. thanks Slavko Bogdanov, Niccolo Bucciantini, Mallory Roberts, and Roger Romani for helpful conversations. P.K. thanks J\"{o}rn Warnecke for helpful discussions. C.V., P.K. \& Z.W. are supported by the South African National Research Foundation (NRF). The work of M.B. is supported by the South African Research Chairs Initiative of the Department of Science and Technology and the NRF\footnote{Any opinion, finding and conclusion or recommendation expressed in this material is that of the authors and the NRF does not accept any liability whatsoever in this regard.}. This work is based on the research supported wholly in part by the NRF (Grant Numbers 87613, 90822, 92860, 93278, and 99072). A.K.H. acknowledges support from the NASA Astrophysics Theory Program. A.K.H., Z.W., and C.V. also acknowledge support from the {\textit{Fermi}} Guest Investigator Cycle 8 Grant. This work has made use of the NASA Astrophysics Data System.

\bibliographystyle{aasjournal}
\bibliography{refs_BWRBs_autoreg}

\end{document}